\documentstyle[11pt,fleqn,epsf,epsfig]{article}
\begin{document}

\title{On Recursive  Production and Evolvabilty of Cells:
Catalytic Reaction Network Approach}

\author{Kunihiko Kaneko\\
{\small \sl Department of Basic Science,
College of Arts and Sciences,}\\
{\small \sl University of Tokyo,}\\
{\small \sl Komaba, Meguro-ku, Tokyo 153, Japan}\\
}

\date{}

\maketitle 

\tableofcontents

\begin{abstract} 
To unveil the logic of cell from a level
of chemical reaction dynamics, we need to clarify how ensemble of
chemicals can autonomously produce the set of chemical, without assuming a
specific external control mechanism.
A cell consists of a huge number of chemical
species that catalyze each other.  Often the number of each molecule
species is not so large, and accordingly the number fluctuations in each molecule species
can be large. In the amidst of such diversity and large fluctuations, how can a cell
make recursive production?  On the other hand, a cell can
change its state to evolve to a different type over a longer time span.  How are reproduction
and evolution compatible?  We address these questions, based on several
model studies with catalytic reaction network.  

In the present survey paper, we first formulate basic questions on the recursiveness and
evolvability of a cell, and then state the standpoint of our research to
answer the questions, that is  termed as 'constructive biology'.  Based on this standpoint, 
we present general strategy of modeling a cell as a chemical reaction network.  

At the first part we investigate of the origin of heredity in a cell, by
noting that the molecules carrying heredity must be preserved well and control
the behavior of a cell.  We take a simpled model consisting of two mutually
catalyzing molecule species, each of which has catalytically active and
inactive types.  One of the molecule species is synthesized slowly, and thus
is a minority in population. Through the growth and division of this cell,
it is shown to reach and remain in a state in which a
 active, minority molecules are preserved over generations, and
control the cell behavior.  This minority controlled state is achieved
by preserving rare number fluctuations of molecules.
The state gives rise to a selection pressure for mechanisms
that ensure the transmission of the minority molecule.  The minority
molecule, thus, carries heredity, and is a candidate for "genetic
information".  Experimental confirmation of this minority control is also
presented.

Next, a protocell model consisting of a large number mutually catalyzing
molecule species is studied, in order to investigate how chemical
compositions are transferred recursively under replication errors.
Depending on the numbers of molecules and species in a cell, and the
path rate in the reaction network, three phases are found: fast
switching state without recursive production, recursive production, and
itinerancy between the above two states. At a recursive production state
chemicals are found to form intermingled hypercycle network that consists of
core hypercycle and peripheral network that influence each other. How
this intermingled network supports the recursive production, and how
minority in the core hypercycle gives rise to a switch to other recursive states
at the itinerancy phase are elucidated.  Evolution of this hypercycle
network is also studied, to show the approach to recursive production of cells and
switch to more efficient reproduction states.  Finally, statistics of
the number distributions of each molecule species are studied,
to show (i)power-law distribution of fast switching
molecules (ii) suppression of fluctuation in the core-network molecule
species and (iii) ubiquity of log-normal distribution for most other
molecule species.  The origin of these statistics are discussed, while
suppression of the number fluctuations of a minority molecule that has
high catalytic connections with others is clarified, that reinforces the
minority control in the replication network.

(Key Words: Minority Control, Heredity, Origin of Life, Constructive Biology
Hypercycle, Chemical Reaction Network, Log-normal Distribution,
Self-reproduction, Evolution)

\end{abstract}

\pagebreak

\section{Basic Question for Recursive Production of a Cell as Reaction Dynamics of Catalytic Network}

{\bf Question: A cell consists of several replicating molecules that
mutually help the synthesis and keep some synchronization for replication.
At least a membrane that partly separates a cell from the outside
has to be synthesized, keeping some degree of synchronization with
the replication of other internal chemicals.  How is such recursive
production maintained, while keeping diversity of chemicals?
Furthermore this recursive production is not complete, and there appears
a slow `mutational' change over generations, which leads to evolution.
How is evolvability compatible with recursive production?\cite{whatlife}}

\subsection{Q1: Origin of Heredity}

In a cell, among many chemicals, only some chemicals (e.g., DNA) are
regarded to carry genetic information.  Why do only some specific
molecules play the role to carry the genetic information?  How has
such separation of roles in molecules between genetic information and
metabolism progressed? Is it a necessary course of a system with
internal degrees and reproduction?

In a cell, however, a variety of chemicals form a complex reaction
network to synthesize themselves.  Then how such cell with a huge
number of components and complex reaction network can sustain
reproduction, keeping similar chemical compositions?

To consider this problem, we start from a simple prototype cell that
consists of mutually catalyzing molecule species whose growth in
number leads to division of the protocell\cite{minority}.  In this
protocell, the molecules that carry the genetic information are not
initially specified.  The first question we discuss here is how
heredity to maintain production of the protocell emerges.  Related
with the question, we ask if there appears some specific molecules to
carry information for heredity, to realize continual reproduction of
such protocell.  We note that in the present cells, it is generally
believed that information is encoded in DNA, which controls the
behavior of a cell.

Here, We do not necessarily take a ``geno-centric" standpoint, in the sense
that gene determines the course of a cell.  In fact, even in these
cells, proteins and DNA both influence their replication process each
other.  Still, it cannot be denied that there exists a difference
between DNA and protein molecules with regards to the role as
information carrier.  In spite of this mutual dependence, why is DNA
molecule usually regarded as the carrier of heredity?
Is there any general rule that some specific molecules play the role of carrier
of genetic information so that the recursive production of cells continues?

Now, the origin of genetic information in a replicating system is an
important theoretical topic that should be studied, not necessarily as
a property of certain molecules, but as a general property of
replicating systems.
To investigate this problem we need to clarify what ''information"  really
means.  In considering information, one often tends to be interested
in how several messages are encoded on a molecule.  In fact, a
hetero-polymer such as DNA would be suited to encode many bits of
information.  One might point out that DNA molecules would be suited
to encode many bits of information, and hence would be selected as an
information carrier.  Although this `combinatorial' capacity of an
information carrier is important, what we are interested here is a
basic property that has to be satisfied prior to that, i.e., origin of
just ``1 bit" information.

As Shannon beautifully demonstrated, information means selection of
one branch from several possibilities \cite{Shannon, Brillouin}.
Assume that there are two possibilities in an event, each of which can
occur with the probability $1/2$.  In this case, when one of these
possibilities turns out to be true, then this choice of a branch is
regarded to have 1 bit information.  In this sense, if a specific cell
state is selected from several possible states, this selection process
has information, and a molecule to control such process carries
information.

Now, a molecule that carries the information is postulated to play the
role to control for the choice of cellular state. Furthermore, to play
the role to carry the information for heredity, the molecules must be
transmitted to next generations relatively faithfully.  These two
features, i.e., control and preservation are nothing but the problem
of heredity.
 
Let us reconsider what 'heredity' really means.  The heredity causes a
high correlation in phenotype between ancestor and offspring.  Then,
for a molecule to carry heredity, we identify the following two
features as necessary.

(1) If this molecule is removed or replaced by a mutant, there is a
strong influence on the behavior of the cell.  We refer to this as the
{\bf `control property'}.

(2) Such molecules are preserved well over generations.  The number of
such molecules exhibits smaller fluctuations than that of other
molecules, and their chemical structure (such as polymer sequence) is
preserved over a long time span, even under potential changes by
fluctuations through the synthesis of these molecules.  We refer to
this as the {\bf `preservation property'}.

These two conditions are regarded as a fundamental condition for a
molecule to establish the heredity.  Now, the problem of `information'
at a minimal level, i.e., 1-bit information is nothing but the problem
of the origin of heredity. As the origin of heredity, we study how a
molecule starts to have the above two properties in a protocell.  In
other words, we study how 1-bit information starts to be encoded on a
single molecule in a replicating cell system.  After we answer this
basic question, we will then discuss how a protocell with the heredity
in the above sense attains incentive to evolve genetic information in
today's sense.

To sum up, the first question we address here is restated as follows.
Consider a protocell with mutually catalyzing molecules.  Then, under
what conditions, recursive production continues maintaining catalytic
activities?  How are recursiveness and diversity in chemicals
compatible?  How is evolvability of such protocells possible?  To
answer these questions, are molecules carrying heredity necessary?
Under what conditions, does one molecule species begin to satisfy the
conditions (1) and (2) so that the molecule carries heredity?  We
show, under rather general conditions in our model of mutually
catalyzing system, that a symmetry breaking between the two kinds of
molecules takes place, and through replication and selection, one kind
of molecule comes to satisfy the conditions (1) and (2).

\subsection{Q2: Recursiveness and Evolvability with Diverse Chemicals}

In a cell, the total number of molecules is limited.  If there are a huge number
of chemical species that catalyze each other, the number of some molecules  species 
may go to zero.  Then molecules that are catalyzed by them no longer are 
synthesized.  Then, other molecules that are catalyzed by them 
cannot be synthesized, either.  In this manner, the chemical compositions 
may vary drastically, and the cell may lose reproduction activity.

Of course, a cell state is not constant, and a cell may not keep on dividing
for ever.  Still, a cell state is sustained to some degree to keep
producing similar offspring cells.  We call such condition for
reproduction of cell as 'recursive production' or 'recursiveness'.
The question we address here is if there are some conditions on
distribution of chemicals or structure of reaction network for recursive production.

There are two directions of study.  One is with regards to the static
aspect of reaction network structure (e.g., topology).  The other is
the number distribution of chemical species and their dynamics.  Of
course, one needs to combine the two aspects to fully understand the
condition for recursive production of a cell.

Currently there are much interest in
the reaction network structure, 
For example, Jeong et al.\cite{Barabasi} studied the metabolic reaction
network, without going into details of the topology.  Write down all
(known) metabolic reaction equations.  Here, the rate of reactions is
disregarded, and only if such reaction equation exists in a cell or
not is concerned.  Then compute how many times a specific molecule
species appears in such reaction equations.  If this number is large,
the molecule species is related with many biochemical reactions.  For
example $H_2O$ has a large number of connections, since in many reactions it appears
either in the left hand or right hand side of the equation.  
$ATP$ has a relatively high number of connections, too.  From these data the
histogram $P(n)$ is obtained, as the number of molecules species that appears $n$
times in the equations.  
From the data, it is shown that
$P(n)$ decays with some power of $n$ as $n^{-\alpha}$\cite{Barabasi}.

So far, the discussion is limited only to topological structure of the
network.  In the reaction network dynamics, the number of molecules
are distributed.  On each 'node' of the network, the abundance of
the corresponding molecule species is assigned.  Accordingly
some path is 'thick' where such reactions occur frequently.  Such
abundance as well as their fluctuations and dynamics has to be
investigated.

In a cell,  the number of each molecule changes in time through
reaction, and the number, on the average is increased for the cell
replication.  For this growth to progress effectively, some positive feedback process
underlying the replication process should exist, which, then, may lead to 
amplification of the number fluctuations in molecules.  With such large 
fluctuations and complexity in
the reaction network, how is recursive production of cells sustained?
Is there any universal statistics in the number distribution of
molecules?

\section{Brief Historical Survey}

\subsection{Eigen's Hypercycle}

Of course, the problem raised in the last section has been
addressed in the study on the origin of life,
or origin of replicating system.  Here we are not necessarily interested in
`what happened in past', but rather, we intend to unveil the universal logic of
cell.  Still, it is relevant to review the earlier studies.

To consider the origin of replication system, one needs to discuss how genetic information
is faithfully transferred to the next generation.
Mills et al.\cite{Spiegelman} set up an experiment of 
RNA replication, by  using a solution of RNA and enzyme.
In this experiment, some enzymes are supplied from outside,
and in this sense it is not an autonomous replication system.
Still, his group found that RNA molecules with proper sequences are
reproduced under some error.

Following this experimental study of Spiegelman on replication of RNA,
Eigen's group started theoretical study on the replication of
molecules\cite{Eigen}.  The replication process of polymer in
biochemical reaction is generally carried out with the aid of enzymes.
The enzyme is given by a polymer, while its catalytic activity
strongly depends on its sequence.  For most sequences of the polymers,
the catalytic activity is very small, but few of them may have high
catalytic activity.  Depending on the sequence some polymer has a much higher
catalytic activity, and the replication rate of polymers depends on the
sequence.  As a theoretical argument, consider replication of
polymers whose replication rate depends on its sequence.  Now, assume
that a 'good' sequence has replication rate $\alpha$ times larger than
its mutant with a substitution of a monomer from the original
sequence.  Here, the replication progresses under some error.  Without
fine machinery for error correction, this error is not negligible.
Assume that in each replication process, a monomer is substituted by
another monomer with the rate $\mu$.  Then the probability that a
polymer consisting of $N$ monomers can produce itself is given by
$(1-\mu)^N \approx exp(-N\mu)$, assuming that $\mu$ is small.

Now, let us examine if the good polymer can continue replication,
maintaining its sequence, so that the information of this
sequence is transferred.  The condition that the good sequence
dominates in populations in the ensemble of polymers is given by

\begin{equation}
N<ln(\alpha)/\mu
\end{equation}

Here,$\ln(\alpha)$ is typically $O(1)$, while the error rate in the
replication of monomer is estimated to be around $0.01\sim0.1$, in
usual polymer replication process.  Then the above condition gives
$N<100$ or so.  In other words, information using a polymer with a
sequence longer than this threshold $N$ is hardly be sustained.  This
problem was first posed by Eigen, and is called 'error
catastrophe'\cite{Eigen}.  On the other hand, information for the
replication for a minimal life system must require much larger
information.  Of course, the error rate could be reduced once some
machinery for faithful replication as in the present life emerges.
However, such machinery requires much more information to be
transmitted by the polymer.

Summing up: For replication to progress, catalysts are necessary, and
information on a polymer to replicate itself must be preserved.  However,
error rate in replication must have been high at a primitive stage of
life, and accordingly, it is recognized that the information to carry
catalytic activity will be lost within few generations.  In other
words, faithful replication system requires larger information, while
a larger information requires faithful replication system.  Thus there
appears catch-22 type paradox.

To resolve this problem of inevitable loss of catalytic activities
through replication errors, Eigen and Schuster proposed
hypercycle\cite{Eigen}, where replicating chemicals catalyze each
other forming a cycle, as ``A catalyzes the synthesis of B, B catalyzes
the synthesis of C, C catalyzes the synthesis of A".  In this case,
each chemical mutually amplifies the synthesis of the corresponding
chemical species in this cycle.  There occurs a variety of mutations
to each species, but this mutant is not generally catalyzed in some
other species in the cycle.  Then, such mutant is not be catalyzed by
C.  This is also understood by writing out the rate equation for the
increase of the population.  In this hypercycle the population
increase is given by the product of the populations of molecules such
as $N_A\times N_B$, $N_B\times N_C$, $N_C \times N_A$, while the
growth of the population of the mutants is linear to each population
$N_A$, $N_B$, $N_C$.  In the previous estimate for error cascade, the
good and mutant sequences increase both linearly to the number.  Then
the the number of variety of mutants dominates.  In the present case,
once the populations of the good sequence in the hypercycle is
dominated, they can sustain the population, against possible emergence
of mutants.  With this hypercycle, the original problem of error
accumulation is avoided.

Since the proposal of hypercycle, population dynamics of molecules for
such catalytic networks have been developed.  However, the hypercycle
itself turned out to be weak against parasitic molecules, i.e., those
which replicate, catalyzed by a molecule in the cycle,
but do not catalyze those in the cycle.  
In contrast to the previous mutant, the growth rate of the
population of these molecules is again the product of the populations
of two species, and such parasitic molecules can invade.

Although the hypercycle itself may be weak against parasitic
molecules, i.e., those which are catalyzed but do not catalyze others,
it is then discussed that compartmentalization by a cell structure may
suppress the invasion of parasitic molecules, or that the
reaction-diffusion system at spatially extended system resolves this
parasite problem\cite{Hogeweg}.  As chemistry of lipid, it is not so surprising
that a compartment structure is formed.  Still, as the origin of life,
this means that more complexity and diversity in chemicals are
required other than a set of information carrying molecules (e.g.,
RNA).

\subsection{Dyson's Loose Reproduction System}

If initially there is a variety of chemicals that form a complex network of
mutual catalyzation, this system may be robust against the invasion of
parasitic molecules.  Such idea resembles stability of ecosystem,
where complex network of several species may resist to invasion of
external species.  Hence we need to study if replication of
complex reaction network can be sustained.  In this case, from the beginning, there
are many molecule species that mutually catalyze, allowing for the existence of
many parasitic molecules.  Here, 
complete replication of the system is probably difficult.  
Then the question we have to
address is if such complex network can maintain molecules that catalyze the synthesis
of the network species.  This question was addressed by Dyson\cite{Dyson},
as a possibility of loose reproduction system.

Dyson, noting the experiment of Oparin on the formation of cell-like
structure, considered a collection of molecules with proteins and
others.  These molecules cannot replicate themselves like DNA or RNA.
They, on the other hand, can have enzyme activities, and catalyze the
synthesis of other molecules albeit not faithful reproduction they may
be.  Still, they may keep similar compositions.  Although
accurate replication of such variety of chemicals is not possible,
chemicals, as a set, may continue reproducing themselves loosely,
while keeping catalytic activity.  Indeed, the accurate replication
must be difficult at the early stage of life, but loose reproduction
could be easier.  However, if this collection of molecules can keep
catalytic activity through reproduction is not evident.

Dyson obtained a condition for the sustainment of catalytic activities
in these collection of molecules, by taking an abstract model.  For
simplicity he classified molecules into two states depending on if
they have catalytic activity or not.  Furthermore, he assumed that the
ratio of the synthesis of catalytic molecules is amplified as the
fraction of catalytic molecules is larger, i.e., a positive feedback
process is assumed.  This model is mapped to a kind of Ising model.
With the aid of mean-field analysis in statistical physics, he showed
that the catalytic activities can be sustained depending on the number
of molecules and their species.  Although his model is abstract, the
result he obtained probably can be applied to any system with a set
of catalytic molecules, be it protein, lipids, or other polymers.

It is important to study if such loose reproduction as a set is
possible in a mutually catalytic reaction network (also see
e.g.\cite{Kauffman,Bagley}).  If this is possible, and if these
chemicals also include molecules forming a membrane for
compartmentalization, reproduction of a primitive cell will become
possible.  In fact, from chemical nature of lipid molecules, it is not
so surprising that a compartment structure is formed.

Still, in this reproduction system, any particular molecules carrying
information for reproduction do not exist, in contrast to the present
cell which has specific molecules (DNA) for it.  As for a transition
from early loose reproduction to later accurate replication with
genetic information, Dyson did not give an explicit answer.
He only referred to 'genetic take-over' that was
originally proposed by Cairns-Smith\cite{Cairns-Smith}, who discussed
that a precise replication system by nucleic acids took over the
original loose reproduction system by clay.  Indeed, Dyson wrote that
his idea is based on `Cairns-Smith theory minus clay'.  However, the
logic for this "take over" is not unveiled.

Considering these theoretical studies so far, it is important to study
how recursive production of a cell is possible, with the appearance of
some molecules to play a specific role for heredity.

\section{Constructive Biology}

\subsection{Standpoint of constructive biology}

Before describing our theoretical model and explaining the numerical
results, it is relevant to briefly summarize our basic standpoint in
the study of biology, termed as "constructive biology"
\cite{whatlife,mtb}\footnote{One can skip this subsection, if one is
not much interested in general standpoint in the study of biology.}.
Here we are interested not in details of specific biological function but
in universal features of a biological
system.  Accordingly we need to study some features that are not
influenced by the details of complicated biological processes.  The
present organisms, however, include detailed elaborated processes that
are captured through the history of evolution.  Then, for our purpose,
it is desirable to set up a minimal biological system, to understand
universal logic that organisms necessarily should obey. Hence, the
approach that should be taken will be 'constructive' in nature.  This
constructive approach is carried out both experimentally and
theoretically.

Our `constructive biology' consists of the following steps of studies.
(i) construct a model system by combining procedures;
(ii) clarify universal class of phenomena through the constructed model(s);
(iii) reveal the universal logic underlying the class of phenomena
and extract logic that the life process should obey;
(iv) provide a new look at data on the present organisms
from our discovered logic.

There are three levels, to perform these steps: 
(1)gedanken experiment ( logic) (2)computer model, and (3)real
experiment.  The first one is theoretical study, reveling a logic
underlying universal features in life processes, essential to
understand the logic of 'what is life'.

Still, life system has a complex relationship among many parts,
which constitute the characteristic feature as a whole, which then influences
the process of each part.
We have not gained sufficient theoretical intuition to
such complex system.  Then it is also relevant to make computer experiments and
heuristically find some logic that cannot be easily reached by logical
reasoning only.  This is the second approach mentioned above, i.e.,
construction of artificial world in a computer.  Here we combine
well-defined simple procedures, to extract a general logic
therein \cite{minority,Complexity,KKTY,Furusawa,speciation}.

Still, in a system with potentially huge degrees of freedom like life,
the construction in a computer may miss some essential factors.
Hence, we need the third experimental approach, i.e., construction in
a laboratory.  In this case again, one constructs a possible biology
world in laboratory, by combining several procedures.  For example,
this experimental constructive biology has been pursued by Yomo and
his collaborators (see e.g., \cite{Matsuura,Ko,Kashiwagi1,Kashiwagi2}
at the levels of biochemical reaction, cell, and ensembles of cells.)

Taking this standpoint of constructive biology,
we have been working problems listed in the table both theoretically and experimentally.
The first two items in the table are related with the construction of
a replicating system with compartment, raised in the questions in \S
1.  Of course, this problem is essential to consider the origin of a
cellular life.  However, we do not intend to reproduce what has
occurred in the earth.  We do not try to guess the environmental
condition of the past earth.  Rather we try to construct such
replication system from complex reaction network under a condition preset
up by us.  For example, by constructing a protocell, in the present
paper, we ask the condition for the heredity, or universal features
of the reaction dynamics to support the recursive production of cells.

The third to sixth items are related with the construction of
multicellular organisms with developmental process.  When cells are
aggregated, they start to form differentiation of roles, and then from
a single cell, robust developmental process to form organized
structure of differentiated cells is generated.  This developmental
process to form a cell aggregate is transferred to the next
generation.  An experimental construction of multi-cellular organisms
(with cell differentiation) from bacteria is one target.  Here again,
we do not try to imitate the process of the present multi-cellular
organisms.  For example, by putting bacteria cells into some
artificial condition, we study if the cells can differentiate into
distinct types or form some robust distribution of cells.  Also,
in-vitro construction of morphogenesis from undifferentiated cells has
been possible by putting cells into some given
conditions\cite{Asashima}. With these studies, we can establish a
viewpoint of universal dynamics underlying development rather than
the conventional picture as finely tuned-up process for it\cite{KKTY,Furusawa}.

The seventh item is construction of evolution, in particular
speciation process, that is how a species splits into two distinct
groups different both in phenotype and genotype\cite{speciation}.

To carry out this plan experimentally we need a system to design a
life system controlled as we like.  Such controlled experiments are
now possible by recent advances in technology, such as flow-cytometry,
imaging techniques, microarray to measure gene expressions, while
advances in nanotechnology provide a powerful tool in constructing a system
to regulate and observe behaviors of a single cell or multiple cells,
in a well controlled situation.

Here this construction is interesting by itself, but our goal is not
the construction itself.  Rather we try to extract general features
that a life system should satisfy, and set up general questions.  For
example, as posed in \S1,  we set up a question if there are some `information molecules'
that control the replication system.  Then we answer the question by
setting up a theory.  For each item, we set up general questions, and
make model simulations, and set up a general theory to answer the
question.  This theoretical part is carried out in tight collaboration
with the experiment.

Table I: examples of constructive biology under current investigation:

\hspace{-.3in}\begin{tabular}{|c||c|c|c|} \hline

construction of &experiment             & theory            & question to be addressed \\ \hline
replicating     & in-vitro replicating  & minority control  & origin of\\
system          & system with several   &                   & information \\
                & enzymes               &                   &                 \\  \hline
cell system     & replicating liposome   & dynamic bottleneck& evolvability \\
                & with internal         & in autocatalytic  & and recursiveness \\
                & reaction network      & reaction system   & for growth       \\ \hline
multicellular   & interaction-induced   & isologous diversi-& robustness in \\
system          & differentiation of    & fication in inter-& development      \\
                & an ensemble of cells  & intra dynamics    &                \\ \hline
developmental   & controlled            & emergence of      & irreversibility  \\
process (I)     & differentiation from  &  differentiation  & in development   \\
                & undifferntiated cells  & rule             &                  \\ \hline
developmental   & activin-controlled    & self-consistency  & origin of      \\
process (II)    & construction of       & between pattern   & positional     \\
                &  tissues formation    &  and dynamics     & information       \\ \hline
generation      & germ-line segregation & higher-level      & origin of recursive \\
                & from ensemble of cells& recursiveness     & individuality      \\ \hline
evolution       & interaction-dependent & symbiotic         & genetic fixation of \\
                & evolution             & sympatric         & phenotypic         \\
                & of E Coli             & speciation        & differentiation    \\ \hline
\end{tabular}

To close this subsection, we give a brief remark on the study of the
so called Artificial Life (AL).  Indeed, our approach may have
something in common with AL\cite{AL}.  In the AL study
people intended to construct life-as-it-could-be, not restricted to
the present organisms.  Originally, in the study of AL, they have been
interested in logic of life that all possible biological system should
obey, be it on this earth or in other conditions in the universe.

Indeed, there are some important studies on the origin of replicating
structure from the side of computation (e.g., \cite{Fontana}).
However, the conventional AL study often tended to imitate life, and could not
propose basic concepts to understand 'what is life'.
Also, the conventional AL study was often biased into the study in
a computer. It often assumes a combination of logical processes with
manipulation of symbols like the study of artificial intelligence.

Our approach is distinct from the conventional artificial
life study in the two points.  First, we do not take such symbol-based
approach, but rather we use dynamical systems approach.  Second, tight
collaboration between experiment and theory is essential.  Note,
however, this collaboration is not of the type to `fit the data' by
some theoretical expression, but rather at a conceptual level.  We
will see an example of such collaboration in \S4.

\subsection{Modeling strategy for the chemical reaction networks}

\begin{figure}
\noindent
\hspace{-.3in}
\epsfig{file=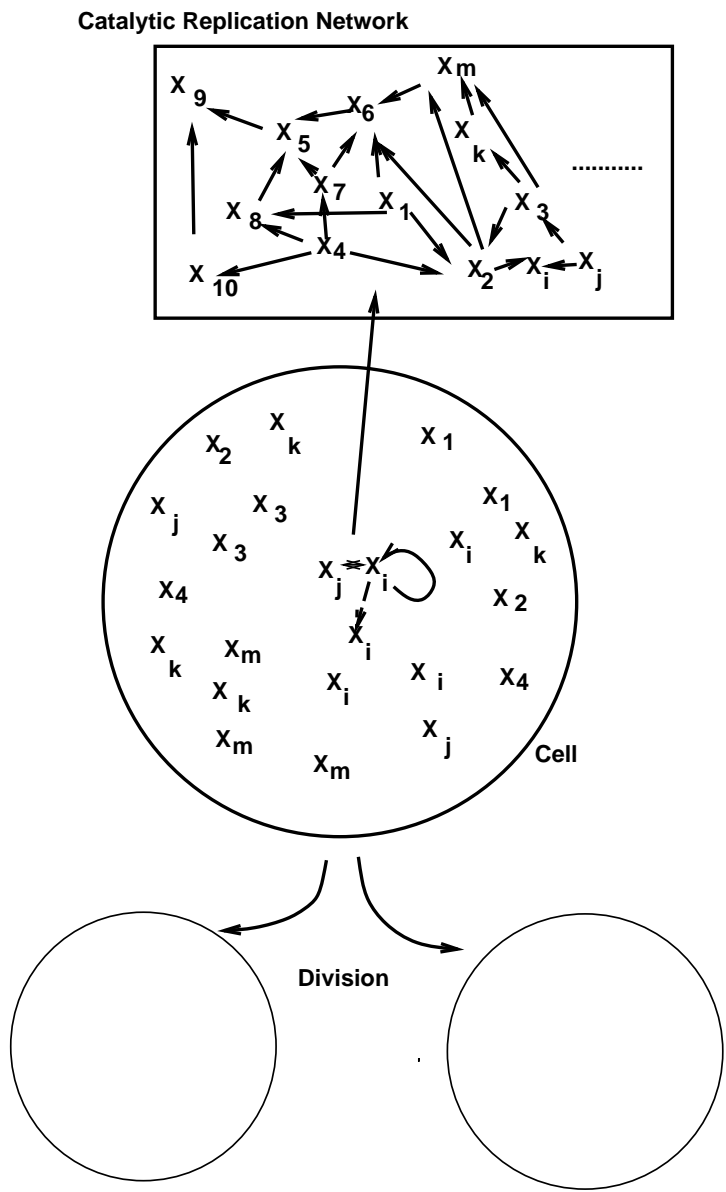,width=.6\textwidth} 
\caption{Schematic representation of our modeling strategy of a cell}
\end{figure}

Now, we discuss a standpoint in modeling cell, based on the standpoint
of the last section.  Then, what type of a model is best suited for a
cell to answer the question in \S 1?  With all the current biochemical
knowledge, we can say that one could write down several types of
intended models.  Due to the complexity of a cell, there is a tendency
of building a complicated model in trying to capture the essence of a
cell.  However, doing so only makes one difficult to extract new
concepts, although simulation of the model may produce similar
phenomena as those in living cells.  Therefore, to avoid such
failures, it may be more appropriate to start with a simple model that
encompasses only the essential factors of living cells.  Simple models
may not produce all the observed natural phenomena, but are
comprehensive enough to bring us new thoughts on the course of events
taken in nature.

In setting up a theoretical model here, we do not put many conditions
to imitate the life process.  Rather we impose the postulates as
minimum as possible, and study universal properties in such system.
For example, as a minimal condition for a cell, we consider a system
consisting of chemicals separated by a membrane.  The chemicals are
synthesized through catalytic reactions, and accordingly the amount of
chemicals increases, including the membrane component.  As the volume
of this system is larger, the surface tension for the membrane can no
longer sustain the system, and it will divide.  After the division of
this protocell systems, they should interact with each other, since
they share resource chemicals.  Under such minimum setup as will be
discussed later, we study the condition for the recursive growth of a
cell, as well as differentiation of the cell.

Let us start from simple argument for a biochemical process that a
cell that grows must at least satisfy.  In a cell, there are a huge
number of chemicals that catalyze each other and form a complex
network.  These molecules are spatially arranged in a cell, and in
some problems such spatial arrangement is very important, while for
some others, the discussion on just the composition of chemicals in a
cell is sufficient to determine a state of a cell.  Hence, for the
starting point we disregard the spatial structure within a cell, and
consider just the composition of chemicals in a cell.  Hence, if there
are $k$ chemical species in a cell, the cell state is characterized by
the number of molecules of each species as $N_1,N_2,...N_k$.  These
molecules change their number through reaction among these molecules.
Since most reactions are catalyzed by some other molecules, the
reaction dynamics consist of a catalytic reaction network.

Through membrane, some chemicals may flow in, which are successively
transformed to other chemicals through this catalytic reaction
network.  For a cell to grow recursively, a set of chemicals has to be
synthesized for the next generation.  As the number of molecules is
large enough, the membrane is no longer sustained, even just due to
the constraint of surface tension.  Then, when the number of molecules
is larger than some value, it is expected to divided.  Hence, the
basic picture for a simple toy cell we take is given as in Fig.1.

Of course, it is impossible to include all possible chemicals in a
model.  As our constructive biology is aimed at neither making
complicated realistic model for a cell, nor imitating specific
cellular function, we set up a minimal model with reaction network, to
answer the questions raised in \S 1.  Now, there are several levels
for the modeling depending on what question we try to answer.

(0) By taking reversible two-body reactions, including all levels of
reactions, ranging from metabolites, proteins, nucleic acids, and so
forth.  For example, to answer the general question, how
non-equilibrium condition is sustained in a cell, such level of model
is desirable\cite{Awazu}.

(1) Assuming that some reaction process are fast, they can be
adiabatically eliminated.  Also, most of fast reversible reactions can
be eliminated by assuming that they are already balanced.
Then we need to discuss only the
concentration (number) of molecules species, that change relatively
slowly.  For example by assuming that enzyme is synthesized and
decomposed fast, the concentrations can be eliminated, to give
catalytic reaction network dynamics consisting of the reactions with

\begin{equation}
X_i+X_j \rightarrow  X_{\ell}+X_j
\end{equation}

\noindent
where $X_j$ catalyzes the reaction\cite{KKTY,Zipf}.  If the catalysis
progresses through several steps, this process is replace by

\begin{equation}
X_i+mX_j \rightarrow  X_{\ell}+mX_j
\end{equation}
leading to higher order catalysis\cite{Furusawa}.  

For a cell to grow, some resource chemicals must be supplied through
membrane.  Through the above catalytic reaction network, the resource
chemicals are transformed to others, and as a result, cell grows.
Indeed, this class of model is adopted to study the condition for cell
growth, to unveil universal statistics for such cells, and also as a
model for cell differentiation. 

(2) Model focusing on the dynamics of replicating units
(e.g.. Hypercycle): For a cell to grow effectively, there should be
some positive feedback process to amplify the number of each molecule
species.  Such positive feedback process leads to autocatalytic
process to synthesize each molecule species. For reproduction of a
cell, (almost) all molecule species are somehow synthesized. Then, it would be possible to take
a replication reaction from  the beginning as a model.  For example, consider a reaction

$S+X+Y \rightarrow X'+Y : S'+X' \rightarrow 2X$. 

\noindent
Then as  a total, the reaction is represented as 

$S+S'+X+Y \rightarrow 2X+Y$. 
\noindent
Assuming the resources S and S' are constantly supplied, we can
consider the replication reaction
\begin{equation}
X+Y \rightarrow 2X+Y,
\end{equation}
catalyzed by $Y$.
At this level, we can take a unit of replicator, and consider a
replication reaction network.  This model was first discussed in the
hypercycle by Eigen and Schuster discussed in \S 2.1.

(3) coarse-grained (phenomenological) level: Some other reduced model
is adopted for the study of gene expression or signal transduction
network.  The modeling at this level is relevant to understand
specific function of a cell.
 
In the present paper we mainly use the modeling of the level (2).
This class of model can be obtained by reducing from the level-(1)
model, by restricting our interest only to take into account of
replicating units.  In this sense, the model is a bit simpler than the
level-(1) model.  On the other hand, it may not be suitable to discuss
the condition for cell growth, since at the level-(2) model, the
supply of resource chemicals is automatically assumed, and one cannot
discuss how transported chemicals are transformed into others.  In the
present paper, we briefly refer to the level-(1) model only  at the end of
\S 5.4, to demonstrate the universality of our result, but for
details, see the original papers \cite{KKTY,Zipf} on the level-(1)
modeling .

To sum up, we envision a (proto)cell containing molecules.  With a
supply of chemicals available to the cell, these molecules replicate
through catalytic reactions, so that their numbers within a cell
increase.  When the total number of molecules exceeds a given
threshold, the cell divides into two, with each daughter cell
inheriting half of the molecules of the mother, chosen randomly.
Regarding the choice of chemical species and the reaction, we discuss
later for specific models. (see Fig.1 for schematic representation).

\section{Minority Control Hypothesis for the Origin of Genetic Information}

In the present section we propose an answer to the question raised in
\S 1.1, by taking a simple model of a cell with replicating molecules,
and proposing a novel concept on minority control, and providing
corresponding experimental results.

\subsection{Model}

As discussed in \S 3.2,
we start from consideration of a prototype of cell, consisting of molecules
that catalyze each other.  As the reaction progresses, the number of
molecules in this protocell will increase.  Then, 
this cell will be divided, when its volume (the total number of molecules)
is beyond some threshold.  Then the molecules split into two `daughter cells".
Then our question in \S 1 is restated as follows:
How are the chemical compositions transferred to the offspring cells?
Do some specific molecules start to carry heredity in the
sense of control and preservation, so that the reproduction continues?

Before considering the specific model,
it may be relevant to recall the difference of roles between DNA (or RNA) and protein.
According to the present understanding of molecular biology\cite{Cell},
changes undergone by DNA molecules are believed to exercise stronger influences
on the behavior of cells than other chemicals.  
Also, a DNA molecule is transferred to offspring cells 
relatively accurately, compared with other constitutes of the cell.
Hence a DNA molecule satisfies (at least) the "preservation" and "control" 
properties (1) and (2) in \S 1.1.

In addition, a DNA molecule is stable, and the time scale for the
change of DNA, e.g., its replication process as well as its
decomposition process, is much slower.  Because of this relatively
slow replication, the number of DNA molecules is smaller than the
number of protein molecules.  At each generation of cells, single
replication of each DNA molecule typically occurs, while other
molecules undergo more replications (and decompositions).

With these natures of DNA in mind, while without assuming the detailed
biochemical properties of DNA, we seek a general condition for the
differentiation of the roles of molecules in a cell and study the
origin of the control and preservation of some specific molecules.

Now, we consider a very simple protocell
system\cite{minority}, consisting of two species of replicating
molecules that catalyze each other (see Fig.2).  assuming that only
two kinds of molecules $X$ and $Y$ exist in this protocell, and they
catalyze each other for the synthesis of the molecules.

\begin{equation}
X + Y \rightarrow 2X+ Y ;Y + X \rightarrow 2Y +X;
\end{equation}

Here, this ``catalytic reaction" is not necessarily a single reaction.
In general there can be several intermediate processes for each
``reaction".  The model simply states that there are two molecules
that help the synthesis of the other, directly or indirectly.  In
general, the catalytic activities as well as the synthesis speeds
differ by types of molecules.  Without losing generality one can
assume that $X$ is synthesized faster than $Y$.

With this synthesis of molecules, the total number of molecules in the
protocell will increase, until it divides into two.  As long as the
molecules catalyze each other, this synthesis continues, as well as
the division (reproduction) of protocell.  However, some structural
changes in molecules can occur through replication (`replication
error').  These structural changes in each kind of molecules may
result in the loss of catalytic activity.  Indeed, the molecules with
catalytic activity are not so common.  On the other hand, molecules
without catalytic activity can grow their number, if they are
catalyzed by other catalytic molecules.  Then, as discussed in \S 2.1,
the maintenance of reproduction is not so easy.

Following the above discussion, we consider the following model,
as a first step in answering the question posed \S 1.1\cite{minority}.

\begin{figure}
\noindent
\hspace{-.3in}
\epsfig{file=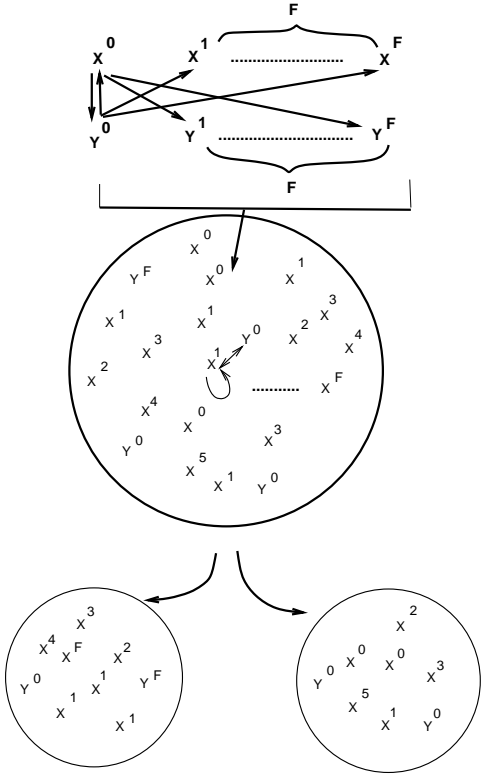,width=.4\textwidth} 
\caption{Schematic representation of our model}
\end{figure}

(i) There are two species of molecules, X and Y, which are mutually catalyzing.

(ii) For each species, there are active and inactive (``I'') types.  
Considering that the active molecule type is rather rare.  There are
$F$ types of inactive molecules per active type.  For most
simulations, we consider the case in which there is only one type of
active molecules for each species.

Active types are denoted as $X^0$ and $Y^0$, while there are inactive
types $X^I$ and $Y^I$ with $I=1,2,...,F$.  The active type has the
ability to catalyze the replication of both types of the other species
of molecules. The catalytic reactions for replication are assumed to
take the form

\begin{math}
X^J + Y^0 \rightarrow 2 X^J +Y^0\end{math} (for $J=0,1,..,F$)

and 
\begin{math}
Y^J + X^0 \rightarrow 2 Y^J +X^0\end{math} (for $J=0,1,..,F$).

(iii) The rates of synthesis (or catalytic activity) of
the molecules $X$ and $Y$ differ.  We stipulate that the rate of the above replication process 
for $Y$,
$\gamma_y$, is much smaller than that for $X$, $\gamma_x$.
This difference in the rates may also be caused by a difference in
catalytic activities between the two molecule species.

(iv) In the replication process, there may occur structural changes
that alter the activity of molecules. Therefore the type (active or
inactive) of a daughter molecule can differ from that of the mother.
The rate of such structural change is given by $\mu$, which is not
necessarily small, due to thermodynamic fluctuations.  This change can
consist of the alternation of a sequence in a polymer or other
conformational change, and may be regarded as replication `error'.
Note that the probability for the loss of activity is $F$ times
greater than for its gain, since there are $F$ times more types of
inactive molecules than active molecules.  Hence, there are processes
described by

\begin{math}
X^I \rightarrow X^0;\end{math}and \begin{math}Y^I \rightarrow Y^0\end{math} (with rate $\mu$)

\begin{math}
X^0 \rightarrow X^I;\end{math}and \begin{math}Y^0 \rightarrow Y^I\end{math}(with rate $\mu $ for each),

resulting from structural change.

(v) When the total number of molecules in a protocell exceeds a given
value $2N$, it divides into two, and the chemicals therein are
distributed into the two daughter cells randomly, with $N$ molecules
going to each.  Subsequently, the total number of molecules in each
daughter cell increases from $N$ to $2N$, at which point these divide.

(vii) To include competition, we assume that there is a constant total
number $M_{tot}$ of protocells, so that one protocell, randomly chosen,
is removed whenever a (different) protocell divides into two.

With the above described process, we have basically four sets of
parameters: the ratio of synthesis rates $\gamma_y/\gamma_x$, the
error rate $\mu$, the fraction of active molecules $1/F$, and the
number of molecules $N$.  (The number $M_{tot}$ is not important, as
long as it is not too small).

We carried out simulation of this model, according to the following procedure.
First, a pair of molecules is chosen randomly. 
If these molecules are of different species, then if the
$X$ molecule is active, a new $Y$ molecule is produced with the probability
 $\gamma_y$, and if the $Y$ molecule is active, a new $X$ 
molecule is produced with the probability $\gamma_x$.
Such replications occur with the error rates given above.
All the simulations were thus carried out
stochastically, in this manner.

We consider a stochastic model rather than the corresponding rate
equation, which is valid for large $N$, since we are interested in the
case with relatively small $N$.  This follows from the fact that in a
cell, often the number of molecules of a given species is not large,
and thus the continuum limit implied in the rate equation approach is
not necessarily justified \cite{Mikhailov}.

Furthermore, it has recently been found that the discrete nature of a
molecule population leads to qualitatively different behavior than in
the continuum case in a simple autocatalytic reaction network
\cite{Togashi}.  In a simple autocatalytic reaction system with a
small number of molecules, a novel steady state is found when the
number of molecules is small, that is not described by a continuum
rate equation of chemical concentrations.  This novel state is first
found by stochastic particle simulations.  The mechanism is now
understood in terms of fluctuation and discreteness in molecular
numbers.  Indeed, some state with extinction of specific molecule
species shows a qualitatively different behavior from that with very
low concentration of the molecule.  This difference leads to a transition to a novel
state, termed as discreteness-induced-transition.  This phase
transition appears by decreasing the system size or flow to the
system, and is analyzed from the stochastic process, where a
single-molecule switch changes the distributions of molecules drastically.

In \cite{Togashi}, given are examples in which a discreteness in molecule
number leads to a novel phase that is not observed from a continuous
rate equation of chemical reaction.  In a cell, since the number of
some molecules species is very small,  we need to seriously consider
the possibility that the discreteness in molecule numbers may lead
to a novel behavior distinct from the continuum description.

\subsection{Result}

If $N$ is very large, the above described stochastic model can be replaced by a
continuous model given by the rate equation.
Let us represent the total number of inactive molecules for each of $X$ and $Y$ as

$ N_x^I =\sum_{j=1}^F N_x^j$; $ N_y^I =\sum_{j=1}^F N_y^j$

Then the growth dynamics of the number of molecules
$N_x^J$ and $N_y^J$
is described by the rate equations, using the total number of molecules $N^t$,

\begin{equation}
dN_x^j/dt=\gamma_x N_x^j N_y^0/N^t;
dN_y^j/dt=\gamma_y N_x^0 N_y^j/N^t.
\end{equation}

From these equations, under repeated divisions,
it is expected that the relations $\frac{N_x^0}{N_y^0}=\frac{\gamma_x}{\gamma_y}$,
$\frac{N_x^0}{N_x^I}= \frac{1}{F}$, and  $\frac{N_y^0}{N_y^I} = \frac{1}{F}$ are eventually satisfied.  
Indeed, even with our stochastic simulation, 
this number distribution is approached as $N$ is increased.

However, when $N$ is small, and with the selection process, there
appears a significant deviation from the above
distribution\cite{minority}.  In Fig.3, we have plotted the average
numbers $\langle N_x^0 \rangle$, $\langle N_x^I \rangle$, $\langle
N_y^0 \rangle$, and $\langle N_y^I \rangle$.  Here, each molecule
number is computed for a cell just prior to the division, when the
total number of molecules is $2N$, while the average $\langle
... \rangle$ is taken over all cells that divided throughout the
simulation.  (Accordingly, a cell removed without division does not
contribute to the average).  As shown in the figure, there appears a
state satisfying $\langle N_y^0 \rangle \approx 2 - 10$, $\langle
N_y^I \rangle \approx 0$.  Since $F \gg 1$, such a state with
$\frac{\langle N_y^0 \rangle}{\langle N_y^I \rangle}>1$ is not
expected from the rate equation (6).  Indeed, for the $X$- species,
the number of inactive molecules is much larger than the number of
active ones.  Hence, we have found a novel state that can be realized
due to the smallness of the number of molecules and the selection
process.


For the dependence of \{$\langle N_x^0 \rangle$,$\langle N_x^I \rangle$,$\langle N_y^0 \rangle$,$\langle N_y^I \rangle$ \}
on these parameters, see also figures of the paper of \cite{minority}.
From these numerical results,
it is shown that the above mentioned state with $\langle N_y^0 \rangle \approx 2 - 10 $, $\langle N_y^I \rangle < 1$
is reached and sustained when $\gamma_y/\gamma_x$ is small and $F$ is sufficiently large.
In fact, for most dividing cells, $N_y^I$ is exactly 0, while there appear a few cells
with $N_y^I>1$ from time to time.
It should be noted that the state with almost no inactive Y
molecules appears in the case of larger $F$, i.e., in the case of
a larger possible variety of inactive molecules.  This suppression of 
$Y^I$ for large $F$ contrasts with the behavior found in the continuum limit (the rate equation).
In Fig.4, we have plotted $\frac{\langle N_y^0 \rangle}{\langle N_y^I \rangle}$ as a
function of $F$.
Up to some value of $F$, the proportion of active $Y$ molecules decreases, 
in agreement with the naive expectation provided by Eq. (6),
but this proportion increases with further increase of $F$,  in the case that
$\gamma_y/\gamma_x$ is small ($\stackrel{<}{\sim}.02$) and $N$ is small.

This behavior of the molecular populations can be understood from the
viewpoint of selection: In a system with mutual catalysis, both $X^0$
and $Y^0$ are necessary for the replication of protocells to continue.
The number of $Y$ molecules is rather small, since their synthesis
speed is much slower than that of $X$ molecules.  Indeed, the fixed
point distribution given by the continuum limit equations possesses a
rather small $N_y^0$.
However, in a system with mutual catalysis, both $X^0$ and $Y^0$ must
be present for replication of protocells to continue.  Note, for the
replication of $X$ molecules to continue, at least a single active $Y$
molecule is necessary.  Hence, if $N_y^0$ vanishes, only the
replication of inactive $Y$ molecules occurs, and divisions from this
cell cannot proceed indefinitely, because the number of $X^0$
molecules is cut in half at each division.  Furthermore, a cell with
$N_y^0=1$, only one of its daughter cells can have an active $Y$
molecule.  Summing up, under the presence of selection, protocells
with $N_y^0>1$ are selected.


On the other hand, the total number of $Y$ molecules is limited to
small values, due to their slow synthesis speed.  This implies that a
cell that suppresses the number of $Y^I$ molecules to be as small as
possible is preferable under selection, so that there is a room for
$Y^0$ molecules.  Hence, a state with almost no $Y^I$ molecules and a
few $Y^0$ molecules, once realized through fluctuations, is expected
to be selected through competition for survival ( see Fig.5 for
schematic representation).  

 Of course, the probability for such rare fluctuations decrease quite
rapidly as the total molecule number increases, and for sufficiently
large numbers, the continuum description of the rate equation is
valid. Clearly then, a state of the type described above is selected
only when the total number of molecules within a protocell is not too
large. In fact, a state with very small $N_Y^I$ appears only if the
total number $N$ is smaller than some threshold value depending on $F$
and $\gamma_y$. In other words, too large cell is not favorable, because
the fluctuation is too small to produce such rare state.

\begin{figure}
\noindent
\hspace{-.3in}
\epsfig{file=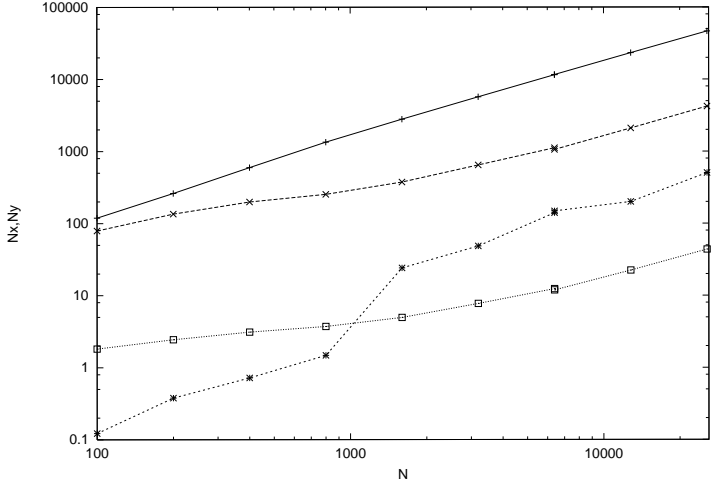,width=.6\textwidth}
\caption{
Dependence of $\langle N_x^0 \rangle (\times)$, $\langle N_x^I \rangle (+)$,
$\langle N_y^0 \rangle (\Box)$, and $\langle N_y^I \rangle(*)$
on $N$.
The parameters were fixed as $\gamma_x=1$, $\gamma_y=0.01$, and $\mu =.05$.
Plotted are the averages of $N_x^0$, $N_x^I$, $N_y^0$, and $N_y^I$
at the division event, and thus their sum is
$2N$.
We use $M_{tot}=100$, and
the sampling for the averages were taken over $10^5-3\times 10^5$ steps,
where the number of divisions ranges from $10^4$ to $10^5$,
depending on the parameters. Reproduced from \cite{minority}.}
\end{figure}

\begin{figure}
\noindent
\hspace{-.3in}
\epsfig{file=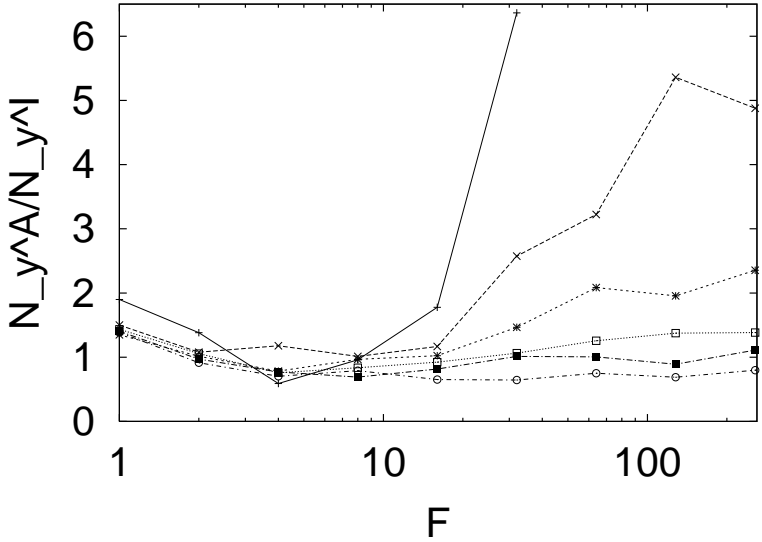,width=.7\textwidth}
\caption{
Dependence of the active-to-inactive ratio, 
$\frac{\langle N_y^0 \rangle }{\langle N_y^I \rangle }$,
on $F$.
The parameters were fixed as $\gamma_x=1$, $\gamma_y=.01$, $\mu =.05$, and $F=128$.
Plots for $\gamma_y=.005$ ($\Diamond$), .01 (+), .015 ($\Box$), 0.02 ($\times$), 
0.025 ($\triangle$),
and 0.03 (*) are overlaid.
Plotted are the averages of $N_x^0$, $N_x^I$, $N_y^0$, and $N_y^I$
at the division event. Reproduced from \cite{minority}.}
\end{figure}


\begin{figure}
\noindent
\hspace{-.3in}
\epsfig{file=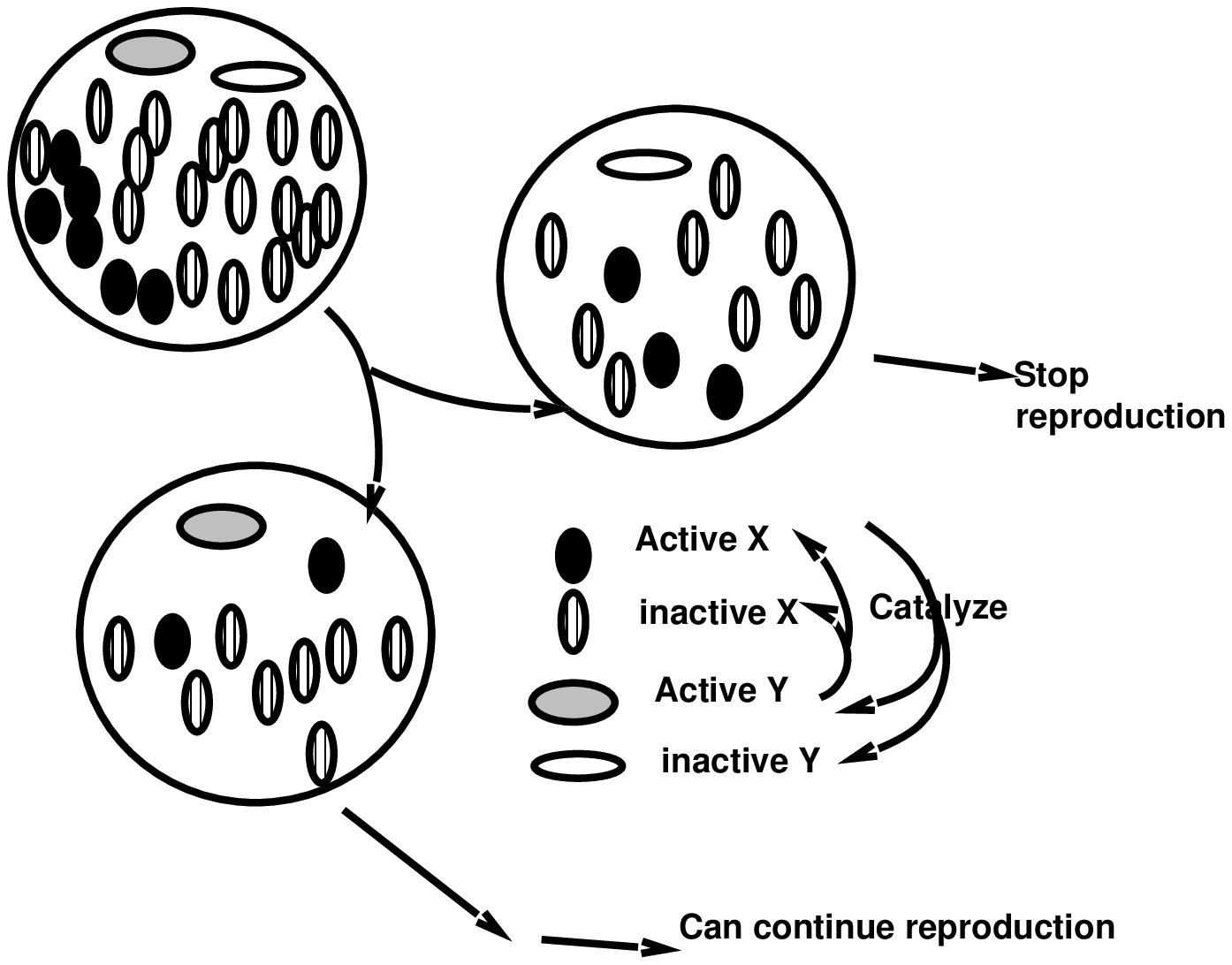,width=.7\textwidth}
\caption{Schematic representation of our logic
Once an active molecule of each molecule species is lost, the
reproduction does not continue.
}
\end{figure}

\subsection{Minority Controlled State}

We showed that in a mutually catalyzing replication system, the
selected state is one in which the number of inactive molecules of the
slower replicating species, $Y$, is drastically suppressed.  In this
section, we first show that the fluctuations of the number of active
$Y$ molecules is smaller than those of active $X$ molecules in this
state.  Next, we show that the molecule species $Y$ (the minority
species) becomes dominant in determining the growth speed of the
protocell system.  Then, considering a model with several active
molecule types, the control of chemical composition through
specificity symmetry breaking is discussed.

\subsubsection{Preservation of minority molecule}

First, we computed the time evolution of the number of active $X$ and
$Y$ molecules, to see if the selection process acts more strongly to
control the number of one or the other.  We computed $N_x^0$ and
$N_y^0$ at every division to obtain the histograms of cells with given
numbers of active molecules.

The fluctuations in the value of $N_y^0$ are found to be much smaller
than those of $N_x^0$.  
The selection process
discriminates more strongly between different concentrations of active
$Y$ molecules than between those of active $X$ molecules.  Hence the
active $Y$ molecules are well preserved with relatively smaller
fluctuations in the number.


\subsubsection{Control of the growth speed}

Now, it is expected that the growth speed of our protocell has a
stronger dependence on the number of active $Y$ molecules than the
number of active $X$ molecules.  We have found that the division time
is a much more rapidly decreasing function of $N_y^0$ than of $N_x^0$.
Even a slight change in the number of active $Y$ molecules has a
strong influence on the division time of the cell.  Of course, the
growth rate also depends on $N_x^0$, but this dependence is much
weaker.  Hence, the growth speed is controlled mainly by the number of
active $Y$ molecules.

\subsubsection{Control of chemical composition by the minority molecule}

As another demonstration of control, we study a model in which there
is more specific catalysis of molecule synthesis.  Here, instead of
single active molecule types for $X$ and $Y$, we consider a system
with $k$ types of active $X$ and $Y$ molecules, $X^{0i}$ and
$Y^{0i}$ ($i=1,2,\cdots k$).  In this model, each active molecule
type catalyzes the synthesis of only a few types ($m<k$) of the other
species of molecules.  Here we assume that both $X$ and $Y$ molecules
have the same ``specificity" (i.e., the same value of $m$) and study
how this symmetry is broken.


As already shown, when $N$, $\gamma_y$ and $F$ satisfy the conditions
necessary for realization of a state in which $N_y^I$ is sufficiently
small, the surviving cell type contains only a few active $Y$
molecules, while the number of inactive ones vanishes or is very
small.  Our simulations show that in the present model with several
active molecule types, only a single type of active $Y$ molecule
remains after a sufficiently long time.  We call this ``surviving
type", $i_r$ ($1 \leq i_r \leq k$).  Contrastingly, at least $m$ types
of $X^0$ species, that can be catalyzed by the remaining $Y^{0i_r}$
molecule species remain.  Accordingly, for a cell that survived after
a sufficiently long time, a single type of $Y^{0i_r}$ molecule catalyzes
the synthesis of (at least) $m$ kinds of $X$ molecule species, while
the multiple types of $X$ molecules catalyze this single type of
$Y^{0i}$ molecules.  Thus, the original symmetry regarding the
catalytic specificity is broken as a result of the difference between
the synthesis speeds.

Due to autocatalytic reactions, there is a tendency for further
increase of the molecules that are in the majority.  This leads to
competition for replication between molecule types of the same
species.  Since the total number of $Y$ molecules is small, this
competition leads to all-or-none behavior for the survival of
molecules. As a result, only a single type of species $Y$ remains,
while for species $X$, the numbers of molecules of different types are
statistically distributed as guaranteed by the uniform replication
error rate.

Although $X$ and $Y$ molecules catalyze each other, a change in the type of
the remaining active $Y$ molecule has a much stronger influence on $X$ 
than a change in the types of the active $X$ molecules on $Y$,
since the number of $Y$ molecules is much smaller.

With the results so far, we can conclude that the $Y$ molecules, i.e.,
the minority species, control the behavior of the system, and are
preserved well over many generations.  We therefore call this state
the minority-controlled (MC) state.

\subsubsection{Evolvability}

An important characteristic  of the MC state is evolvability.
Consider a variety of active molecules $0i$, with different catalytic activities.
Then the synthesis rates $\gamma_x$ and $\gamma_y$ depend on the activities of
the catalyzing molecules.  Thus, $\gamma_x$ can be written in terms of
the molecule's inherent growth rate, $g_x$, and the activity, $e_y(i)$, of
the corresponding catalyzing molecule $Y^{0i}$:

\begin{math}
\gamma_x =g_x \times e_y(i);
\gamma_y =g_y \times e_x(i).
\end{math}

\noindent
Since such a biochemical reaction is entirely facilitated by catalytic
activity, a change of $e_y$ or $e_x$, for example by the structural
change of polymers, is more important. Given the occurrence of
such a change to molecules, those with greater catalytic activities
will be selected through competition evolution, leading to the
selection of larger $e_y$ and $e_x$.  As an example to demonstrate
this point, we have extended the model to include $k$ kinds of active
molecules with different catalytic activities.  Then, molecules with
greater catalytic activities are selected through competition.

Since only a few molecules of the $Y$ species exist in the MC state, a
structural change to them strongly influences the catalytic activity
of the protocell.  On the other hand, a change to $X$ molecules has a
weaker influence, on the average, since the deviation of the {\sl
average} catalytic activity caused by such a change is smaller, as can
be deduced from the law of large numbers.  Hence the MC state is
important for a protocell to realize evolvability.

\subsection{Experiment}

Recently, there have been some experiments to construct minimal
replicating systems in vitro.  
As an experiment corresponding to this problem, we describe an in-vitro
replication system, constructed by Yomo's group\cite{Matsuura}. 

In general, proteins are synthesized from the information on DNA
through RNA, while DNA are synthesized through the action of proteins.
As a set of chemicals, they autonomously replicate themselves.  Now
simplifying this replication process, Matsuura et al.\cite{Matsuura} constructed a
replication system consisting of DNA and DNA polymerase i.e., an
enzyme for the synthesis of DNA, and so forth.  This DNA polymerase is
synthesized by the corresponding gene in the DNA, while it works as
the catalyst for the corresponding DNA.  Through this mutual catalytic
process the chemicals replicate themselves.

\begin{figure}
\noindent
\hspace{-.3in}
\epsfig{file=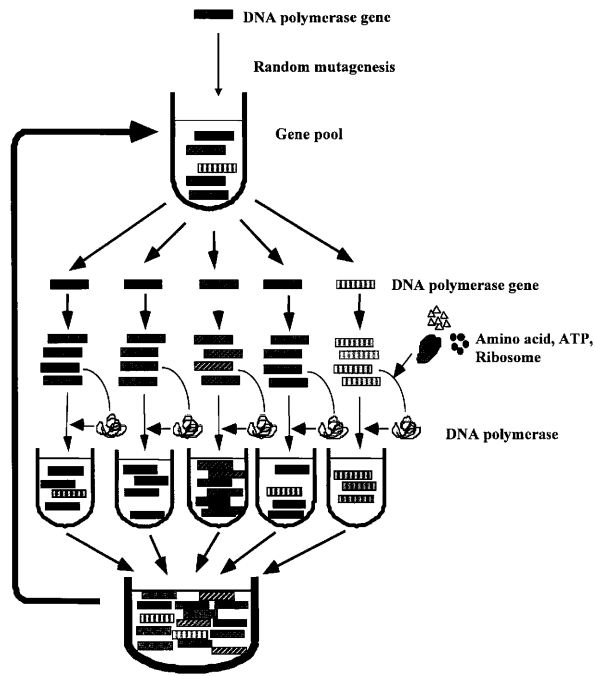,width=.9\textwidth}
\caption{
Illustration of in-vitro autonomous replication system
consisting of DNA and DNA polymerase. 
See text and \cite{Matsuura} for details.
Provided with the courtesy of Yomo, Matsuura et al.
}
\end{figure}




\begin{figure}
\noindent
\hspace{-.3in}
\epsfig{file=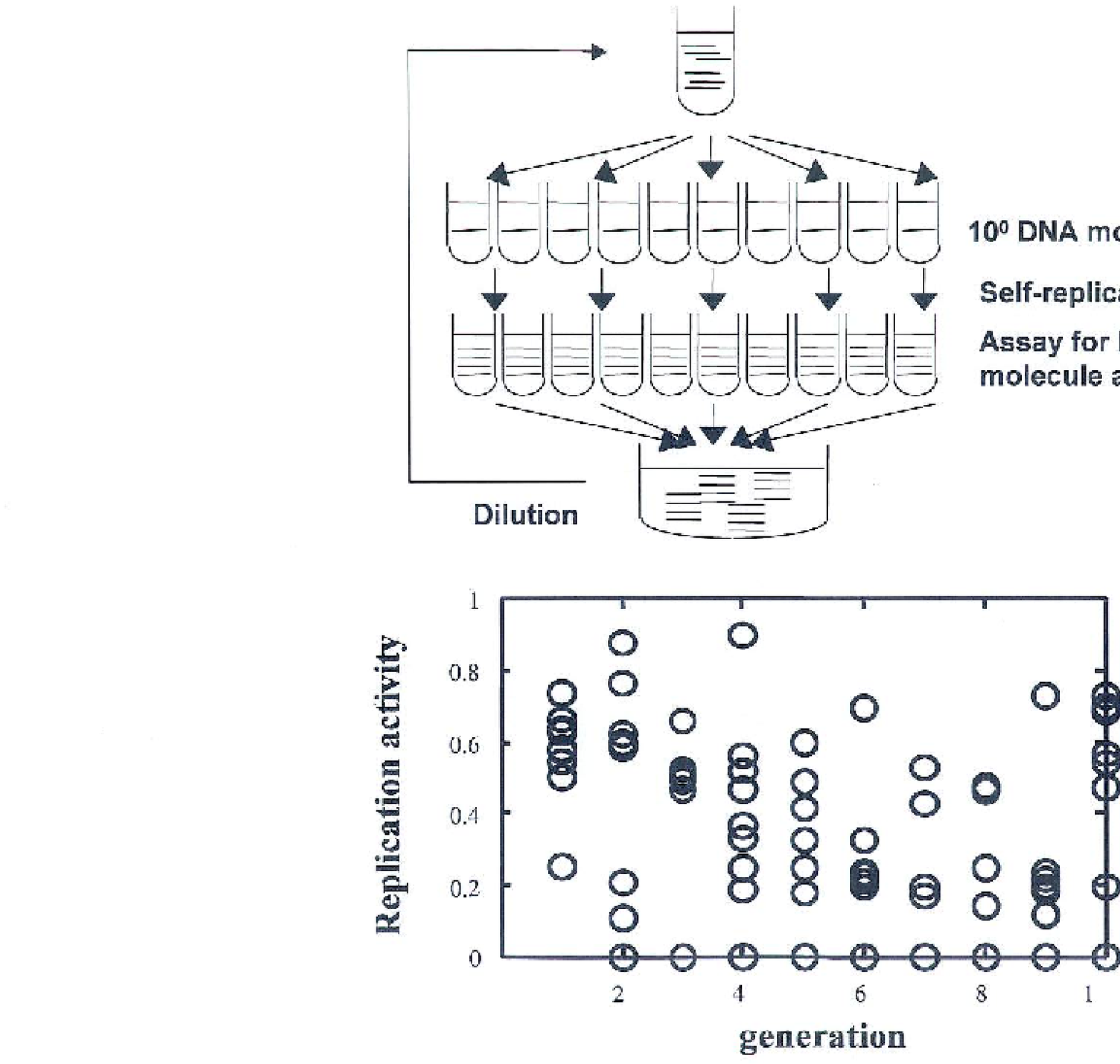,width=.7\textwidth}
\epsfig{file=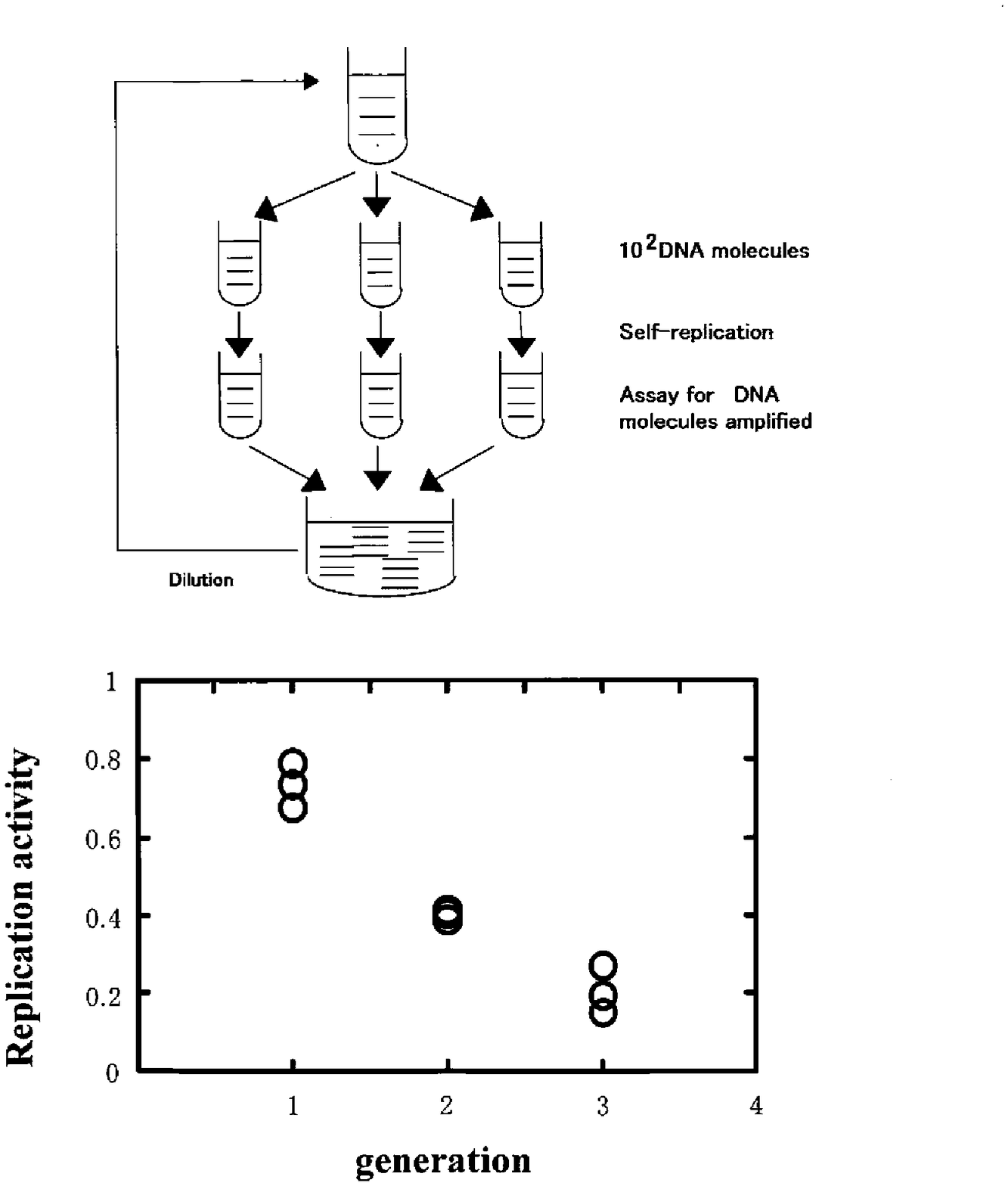,width=.7\textwidth}
\caption{ 
Self-replication activities for each generation, measured as described in the
text. The activities for 10 tubes are shown.  :  Upper: result from a single DNA, where the  next generation is
produced mostly from the top DNA.  Although activities vary by each tube,
higher ones are  selected, so that the activities are maintained. Lower: result from 100 DNA molecules.
Provided with the courtesy of
Yomo, Matsuura et al\cite{Matsuura}.
}
\end{figure}

As for the amplification of DNA, PCR is widely used, and is a
standard tool for molecular biology.  In this case, however, enzymes
that are necessary for the replication of DNA must be supplied
externally.  In this sense, it is not a self-contained autonomous
replication system.  In the experiment by Yomo's group, while they use
PCR as one step of experimental procedures, the enzyme (DNA
polymerase) for DNA synthesis is also replicated in vitro within the
system.  Of course, some (raw) material, such as amino acid or ATP,
have to be supplied, but otherwise the chemicals are replicated by
themselves. (see Fig. 6 for the experimental procedure).

In this experiment, there is mutual synthetic process between gene and enzymes.
Roughly speaking, the polymerase in the experiment corresponds to
$X$ in our model, while the polymerase gene corresponds to $Y$.

Now, at each step of replication , about $2^{30}\sim2^{40}$ DNA molecules are replicated.
Here, of course there are some errors. These errors can occur in the synthesis of
enzyme, and also in the synthesis of DNA.  With these errors, there appear DNA molecules
with different sequences.  Now a pool of DNA molecules with a variety of sequences 
is obtained as a first generation.

From this pool, the DNA and enzymes are split into several tubes.
Then, materials with ATP and amino acids are supplied, and the replication process
is repeated (see Fig. 6).  In other words, the 'test tube' here plays the role
of ``cell compartmentalization".  Instead of autonomous cell division, split into several tubes are operated externally.  

In this experiment, instead of changing the synthesis speed $\gamma_y$ or $N$ in the model, 
one can control the number of genes, by changing the condition how
the pool is split into several test tubes.

Indeed, they studied the two distinct cases, i.e., 
split to tubes containing a single DNA in each and split 
to tubes containing 100 DNA molecules.
Recall that  in the theory, the evolvability by minority control is predicted.
Hence, the behavior between the two cases may be drastically different.

First, we describe the case with a single DNA in each tube.
Here, the pool of chemicals 
is split into 10 tubes each of which has a single DNA molecule, 
and replication process described already
progresses in each tube.  Here, the sequence of DNA molecules could be different by tube,
since there is replication error.  Then the activity of DNA polymerase by each
tube is also different, and the number of DNA molecules synthesized in each tube
is different.  In other words,  some DNA molecules can produce more offspring, but others
cannot.  The variation of self-replication activity by tubes is shown in the upper column of Fig. 7.
Then the contents of each tube are mixed.
This soup  of chemicals is used for the next generation.  Then in this soup, 
the DNA molecules that have higher replication rate as well as  their mutants generated
from them are included with a larger fraction.  Now a single DNA is selected from
the soup in each of 10 tubes, and the same procedures are repeated.
Hence, there is a larger probability that a DNA molecule with a
higher reproduction activity is selected for the next generation. In other words,
Darwinian selection acts at this stage.
The self-replication activity
from this soup is plotted in the third generation.  Successive plots of the 
self-replication activity are given in the upper column of Fig. 7,
As shown, the self-replication activity is not lost (or can evolve in some case),
although it varies by each tube in each generation.

One might say that the maintenance of replication is not surprising at all,
since a gene for the DNA polymerase is included in the beginning.
However, enzyme with such  catalytic activity is rare. Indeed, with mutations
some proteins that lost such catalytic activity but are synthesized in the
present system could appear, which might take over the system.
Then the self-replication activity would be lost.  In fact, this is nothing but 
the error catastrophe by Eigen, discussed in \S 2.1.
Then, why is the self-replication activity maintained in the present experiment?

The answer is clear according to the theory in \S 4.2-4.3.
In the model of \S 4.1, mutants that lost the catalytic
activity are much more common(i.e., $F$ times larger in the model).
Still, the number of such molecules is suppressed.  This was possible
first because the molecules are in a cell.  In the experiment also
they are in a test tube, i.e., in a compartment.  Now the selection works
for this compartment, not for each molecule.  Hence the tube (cell) that
includes a gene giving rise to lower enzyme activity produces less offspring.
In this sense, compartmentalization is one essential factor for
the maintenance of catalytic activity (see also \cite{Hogeweg,Szathmary,Eigen-book}.
Here, another important factor is that in each compartment (cell) there is a single (or
very few)  DNA molecule (as the $Y$ molecule in the model of \S 4.1-3).  In the theory,
if the number of $Y$ molecules is larger,  inactive $Y$ molecules surpass
the active one in population.

To confirm the validity of our theory, Matsuura et al.\cite{Matsuura} carried out a comparison
experiment.  Now, they split the chemicals in the soup so that each tube
has 100 DNA molecules instead of a single one.  Otherwise, they adopt
the same procedure.  In other words, this corresponds to a cell with 100
copies of genome.  Change of self-replication activity in the experiment
is plotted in the lower column of Fig.7.  As shown, the
self-replication activity is lost by each generation, and after the
fourth generation, capability of autonomous replication is totally lost.
This result shows that the number of molecules to carry genetic
information should be small, which is consistent with the theory.

When there are many DNA molecules, there can be mutation to each DNA
molecule.  In each tube, the self-replication activity is given by the
average of the enzyme activities from these 100 DNA molecules.
Although catalytic activity of molecules varies by each, the variance
of the average by tubes should be reduced drastically.  Recall that
the variance of the average of $N$ variables with the variance $\mu$
is reduced to $\mu/N$, according to the central limit theorem of
probability theory.  Hence the average catalytic activity does not
differ much by tube.  Here, the mutant with a higher catalytic
activity is rare.  Most changes in the gene lead to smaller or null
catalytic activity.  Hence, on the average, the catalytic activity
after mutations to original gene gets smaller, and the variance by
tubes around this mean is rather small (see Fig. 7).

By the selection, DNA from a tube with a higher catalytic activity
could be selected, but the variation by tubes is so small that the
selection does not work.  Hence deleterious mutations remain in the
soup, and the self-replication activity will be lost by generations.
In other words, the selection works because the number of information
carrier in a replication unit (cell) is very small, and is free from
the statistical law of large numbers.

Summing up: In the experiment, it was found that replication is
maintained even under deleterious mutations (that correspond to
structural changes from active to inactive molecules in the model),
only when the population of DNA polymerase genes is small and
competition of replicating systems is applied.  When the number of
genes (corresponding to $Y$) is small, the information containing in
the DNA polymerase genes is preserved.  This is made possible by the
maintenance of rare fluctuations, as found in our theory.  The system
has evolvability only if the number of DNA in the system is small.
Otherwise, the system gradually loses its activity to replicate
itself.  These experimental results are consistent with the minority
control theory described.

\subsection{Discussion}

\subsubsection{Heredity from a kinetic viewpoint}

In this section, we have shown that in a mutually catalyzing system,
molecules $Y$ with the slower synthesis speed and minority in number,
tend to act as the carrier of heredity. Through the selection under
reproduction, a state, in which there is a few active $Y$ and almost zero
inactive $Y$ molecules, is selected.  This state is termed the
``minority controlled state".  Between the two molecule species, there
appears separation of roles, between that with a larger number, and that with
a greater catalytic activity.  The former has a variety of chemicals
and reaction paths, while the latter works as a basis for the
heredity, in the sense of the two properties mentioned in \S 1.1 and
\S 4.3, `preservation' and `control'.  We now discuss these properties
in more detail.

[Preservation property]: A state that can be reached only through
very rare fluctuations is selected, and
it is preserved over many generations, even though
the realization of such a state is very rare
when we consider the rate equation obtained in the continuum limit.

[Control property]: A change in the number of $Y$ molecules 
has a stronger influence on the growth rate of a cell than a change
in the number of $X$ molecules.
Also, a change in the catalytic activity of the $Y$ molecules has a strong 
influence on the growth of the cell.  The catalytic activity of the $Y$ 
molecules acts as a control parameter of the system. 

Once this minority controlled state is established, the following
scenario for the evolution of genetic information is expected.  First,
a new selection pressure is now possible to emerge, to evolve a
machinery to ensure that the minority molecule makes it into the
offspring cells, since otherwise the reproduction of the cell is
highly damaged.  Hence a machinery to guarantee the faithful
transmission of the minority molecule should evolve.  Now, the origin
of heredity is established.  Here, for this heredity, any specific
metabolic or genetic contents transmitted faithfully is not necessary.
It can appear from the loose reproduction system that Dyson considered
(as in \S 2.2).  This heredity evolves just as a result of kinetic
phenomenon and is a rather general phenomenon in a reproducing
protocell consisting of mutually catalytic molecules.

This faithful transmission of minority molecule provides a basis for
critical information for reproduction of the protocells.  Since this
minority molecule is protected to be transmitted, other chemicals that
are synthesized in connection with it are probable to be transmitted,
albeit not always faithfully.  Hence there appears a further
evolutionary incentive to package life-critical information into the
minority molecule.  Now more information (`many bits' of information)
are encoded on the minority molecule.  Then, the molecules work as a
carrier of genetic information in the today's sense.  With this
evolution having more molecules catalyzed by the minority molecule, it
is then easier to further develop the machinery to better take care of
minority molecules, since this minority molecule is essential to many
reactions for the synthesis of many other molecules.

Hence the evolution of faithful transmission of minority molecules and
of coding of more information reinforce each other.  At this point one
can expect a separation of metabolism and genetic information.

To sum up, how a single molecule starts to reign the heredity is
understood from a kinetic viewpoint.  We first show the minority
controlled state as a rather general consequence of kinetic process of
mutually catalytic molecules. This provides a basis for heredity.
Taking advantage of the evolvability of minority controlled state,
then, preservation mechanism of the minority molecule evolves, which
allows for more information encoded on it, leading to separation of
genetic information and metabolism.  In this sense, the minority
molecule species with slower synthesis speed, leading to the
preservation of rare states and control of the behavior of the system,
acts as an information carrier. The important point of our theory is
that heredity arises prior to any metabolic information that needs to
be inherited.






\subsubsection{some remarks}

In \S 2, we described two standpoints on the origin of life, i.e.,
genetic information first or complex metabolism first.  We pointed out
some difficulty at each standpoint.  In the former picture, there was
a problem on the stability against parasites, while the latter cannot
solve how genetic information took over the original loose
reproduction system.  The minority control gives a new look to these
problems.

The first problem in \S 2.1  was the appearance of parasitic molecules to destroy
the hypercycle, i.e. mutually catalytic reaction cycle.  If only the
replication process of molecules is concerned, it is not so easy to
resolve the problem.  Here we consider the dual level of replication,
i.e., molecular and cellular replication.

In the present theory for the origin of information, existence of a
cell unit that reproduces itself is required.  Two levels of
reproduction, both molecules and cells are assumed here.  Hence a cell
with parasitic molecules cannot grow, and is selected out.  Relevance
of this type of two-level reproduction to avoid molecular parasites
has been discussed \cite{Hogeweg,Szathmary,Eigen-book}.  Here,
relevance of cellular compartment to the {\em origin of genetic
information} is more important.

This two-level selection works effectively, with the aid of minority
control of specific molecules for a cell.  Indeed, surviving cells
satisfy the minority control.  With the selection pressure for
reproduction of cells, there appears a state that is not expected by
the rate equation for reaction of molecules, where the number of
inactive $Y$ molecules that are parasitic to the catalytic reaction is
suppressed.  Furthermore, resistance against parasitic (inactive) $Y$
molecules is established by this minority controlled state.

This minority control also resolves the question on the genetic
take-over, the problem in the ''metabolism first'' standpoint (in \S 2.2).  Among
several molecules, specific molecule species that are minority in
population controls the behavior of a cell and is well preserved.  The
possible scenario mentioned in the beginning of this section gives one
plausible answer how genetic take-over progresses.


The differentiation of role between the molecules looks like
``symmetry breaking''.  When initially two states are equally
possible, and later only one of them is selected, it is said that the
symmetry is broken. In the differentiation of roles of molecules
studied here, however, the molecules have different characters as to
the replication speed from the beginning. Here a difference in one
character (i.e., the replication speed) is ''transformed'' into the
difference in the control behavior, and in the role as a carrier
of heredity.  In other words, a characteristics with already broken
symmetry is transformed into a different type of symmetry breaking.
This kind of transformation of one character's difference to another
is often seen in biology, as we have already discussed in the study of
morphogenesis and sympatric speciation\cite{Furusawa,speciation}.

\section{Recursive Production in an Autocatalytic Network}

Now we come to the second question raised in \S 1.  In the model of
the last section, we considered a system consisting of two kinds of
molecules.  In a cell, however, a variety of chemicals form a complex
reaction network to synthesize themselves.  Here we study a model with
a large number of chemical species, to discuss how a cell with such
large number of components and complex reaction network can sustain
reproduction, keeping similar chemical compositions
\cite{KK-net,KK-PRE}(see also \cite{Lancet}).

\subsection{Model}

To unveil general features of a system with mutually catalyzing
molecules, we study a system with a variety of chemicals ($k$ molecule
species), forming a mutually catalyzing network.  The molecules
replicate through catalytic reactions, so that their numbers within a
cell increase.  (see Fig.1 again for schematic representation of the
model).

We envision a (proto)cell containing $k$ molecular species with some
of the species possibly having a zero population.  A chemical species
can catalyze the synthesis of some other chemical species as

\begin{equation}
[i] + [j] \rightarrow [i] + 2[j],
\label{reaction}
\end{equation}

\noindent
with $i,j=1,\cdots,k$ according to a randomly chosen reaction network,
where the reaction is set at far-from-equilibrium, In eq.(7), the
molecule $i$ works as a catalyst for the synthesis of the molecule
$j$, while the reverse reaction is neglected, as discussed in the
hypercycle model.  For each chemical the rate for the path of
catalytic reaction in eq.(7) is given by $\rho$, i.e., each species has
about $k\rho$ possible reactions.  The rate is kept fixed throughout
each simulation.  Considering catalytic reaction dynamics, the reverse
reaction process is neglected, and reactions $i \leftrightarrow j$ are
not included.  (Here we investigated the case without direct mutual
connections, i.e., $i\rightarrow j$ was excluded as a possibility when
there was a path $j \rightarrow i$, although this condition is not
essential for the results to be discussed).  Furthermore, each
molecular species $i$ has a randomly chosen catalytic ability $c_i \in
[0,1]$ (i.e., the above reaction occurs with the
rate $c_i$).   Assuming an environment with an ample
supply of chemicals available to the cell, the molecules then
replicate leading to an increase in their numbers within a cell.

Again, when the total number of molecules exceeds a given threshold
(here we used 2$N$), the cell is assumed to divide into two, with each
daughter cell inheriting half of the molecules of the mother cell,
chosen randomly.

During the replication process, structural changes, e.g., the
alternation of a sequence in a polymer, may occur that alter the
catalytic activities of the molecules.  Therefore, the activities of
the replicated molecule species can differ from those of the mother
species.  The rate of such structural changes is given by the
replication 'error rate' $\mu$.  As a simplest case, we assume that
this `error' leads to all other molecule species with equal
probability (i.e., with the rate $\mu /(k-1)$),  and could thus
regard it as a background fluctuation.  In reality, of course, even
after a structural change, the replicated molecule will keep some
similarity with the original molecule, and a replicated species with
the `error' would be within a limited class of molecule species.
Hence, this equal rate of transition to other molecule species is a
drastic simplification.  Some simulations where the errors in
replication only lead to a limited range of molecule species, however,
show that the simplification does not affect the basic conclusions
presented here.  Hence we use the simplest case for most simulations.

In statistical physics, people study mostly the case the total number
of molecules $N$ is very large, at least much much larger than a
number of molecule species $k$.  In this case, the continuum
description is relevant.  When $N/k$ is rather small, some molecules
species can often fluctuate around 0, where the discreteness
0,1,2,... will be important, as already discussed.  In order to take
the importance of the discreteness in the molecule numbers into
account, we adopted a stochastic rather than the usual differential
equations approach, by taking a variety of possible chemicals, where
$N$ and $k$ are of a comparable order.

The model is simulated as follows: At each step, a pair of molecules,
say, $i$ and $j$, is chosen randomly.  If there is a reaction path
between species $i$ and $j$, and $i$ ($j$) catalyzes $j$ ($i$), one
molecule of the species $j$ ($i$) is added with probability $c_i$
($c_j$), respectively.  The molecule is then changed to another
randomly chosen species with the probability of the replication error
rate $\mu$.  When the total number of molecules exceeds a given
threshold (denoted as $N$), the cell divides into two such that each
daughter cell inherits half ($N/2$) of the molecules of the mother
cell, chosen randomly\cite{minority}.

Again, to include competition, we assume that there is a constant
total number $M_{tot}$ of protocells, so that one protocell, randomly
chosen, is removed whenever a (different) protocell divides into two.
However, the result here does not depend on $M_{tot}$ so much.  We
choose mostly $M_{tot}=1$, in the results below but the simulation
with $M_{tot}=100$ gives essentially the same behavior.




\subsection{Result}

\subsubsection{Phases}

Our main concern here is the dynamics of these molecule numbers $N_i$
of the species $i$ in relationship with the condition of the recursive
growth of the (proto)cell.  In our model there are four basic
parameters; the total number of molecules $N$, the total number of
molecule species $k$, the mutation rate $\mu$, and the reaction path
rate $\rho$.  By carrying out simulations of this model, choosing a
variety of parameter values $N,k,\mu,\rho$, also by taking various
random networks, we have found that the behaviors are classified into
the following three phases\cite{KK-net,KK-PRE}:

(1) Fast switching states without recursiveness

(2) Achievement of recursive production  with similar chemical compositions

(3) Switch over several quasi-recursive states

\begin{figure}  
\noindent  
\epsfig{file=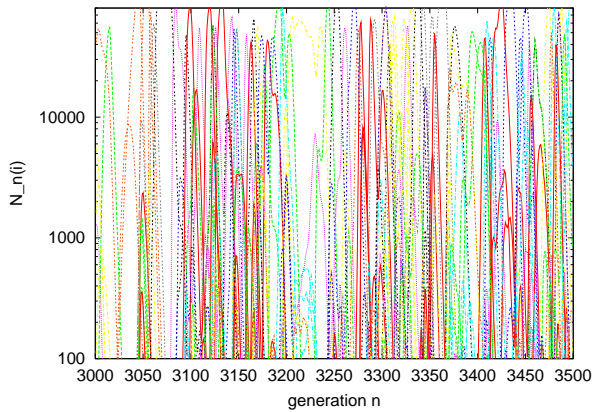,width=.5\textwidth}  
\caption{The number of molecules $N_n(i)$ for the species $i$ is
plotted as a function of generation $n$ of cells, i.e., at each
successive division event $n$.  A random network with $k=500$ and
$\rho=.2$.  Dominant species change successively in generation.}
\end{figure}

\begin{figure}
\noindent
(a)\epsfig{file=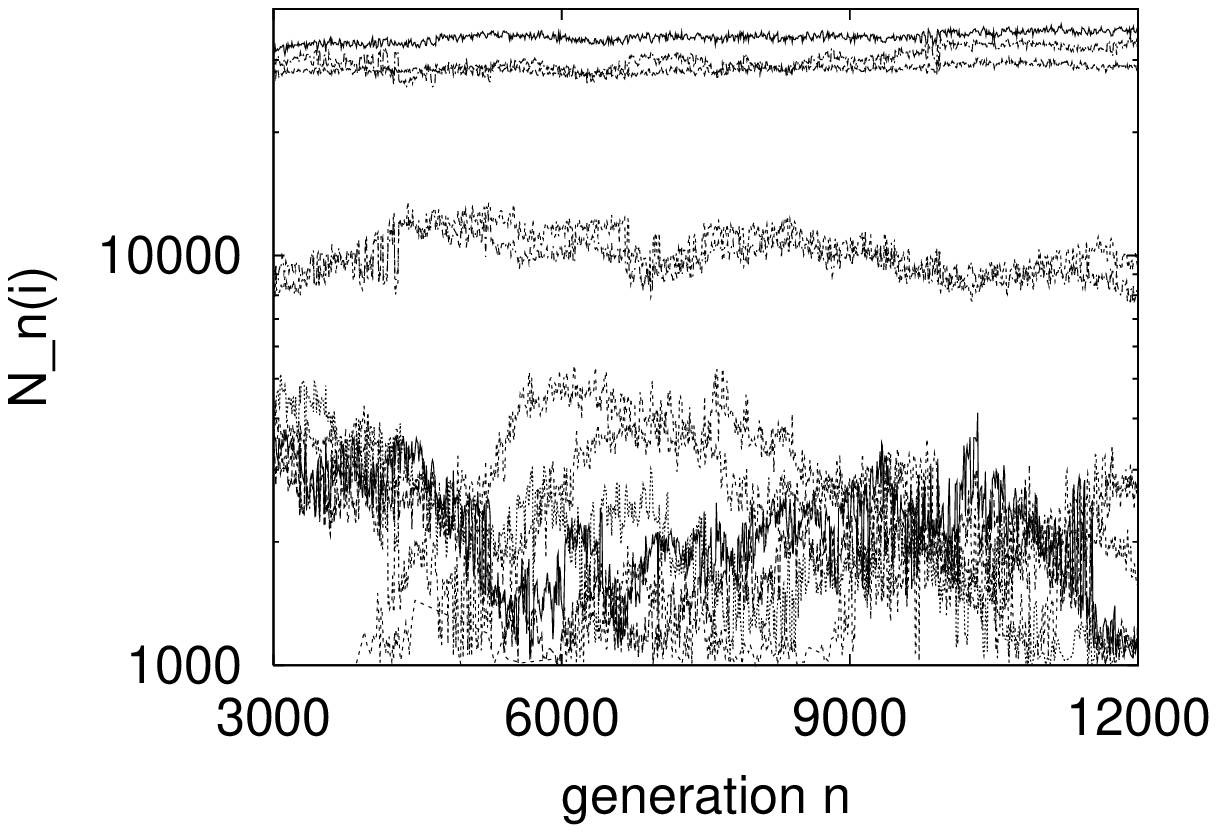,width=.53\textwidth}
(b)\epsfig{file=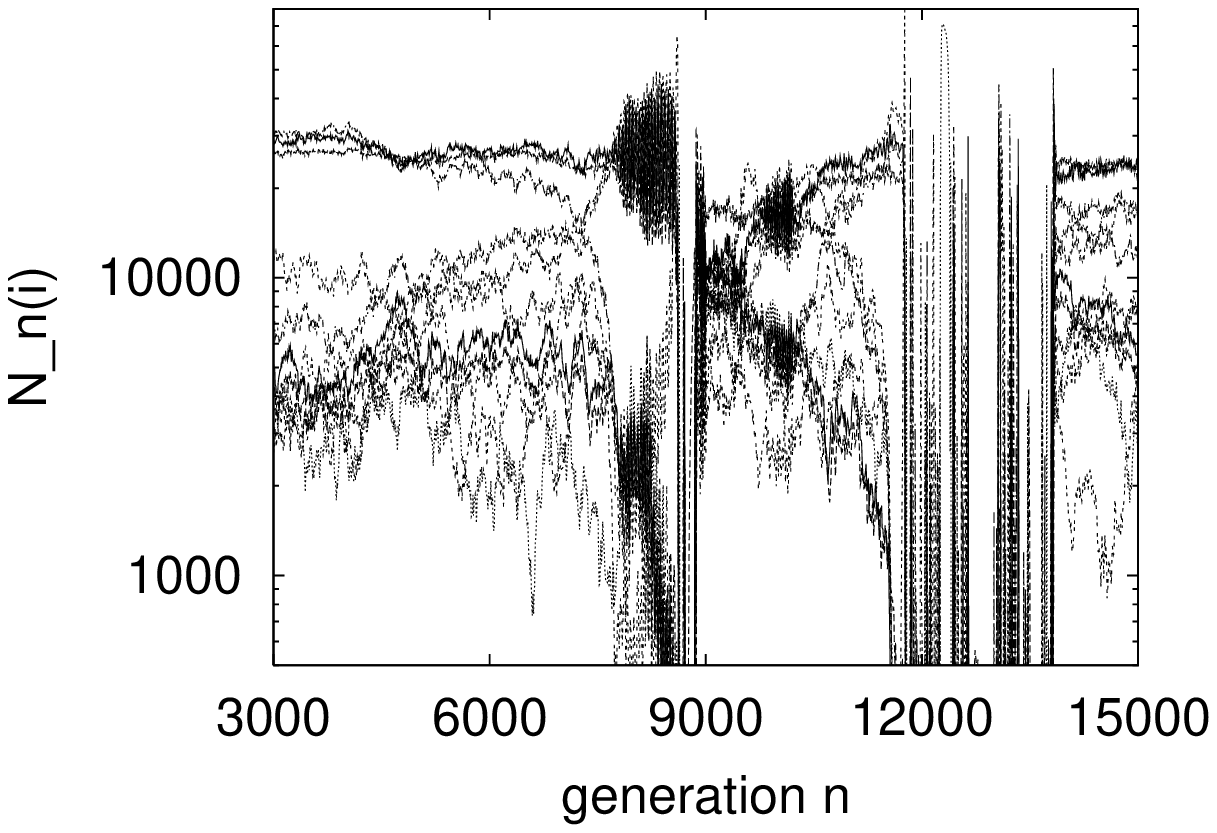,width=.53\textwidth}
(c)\epsfig{file=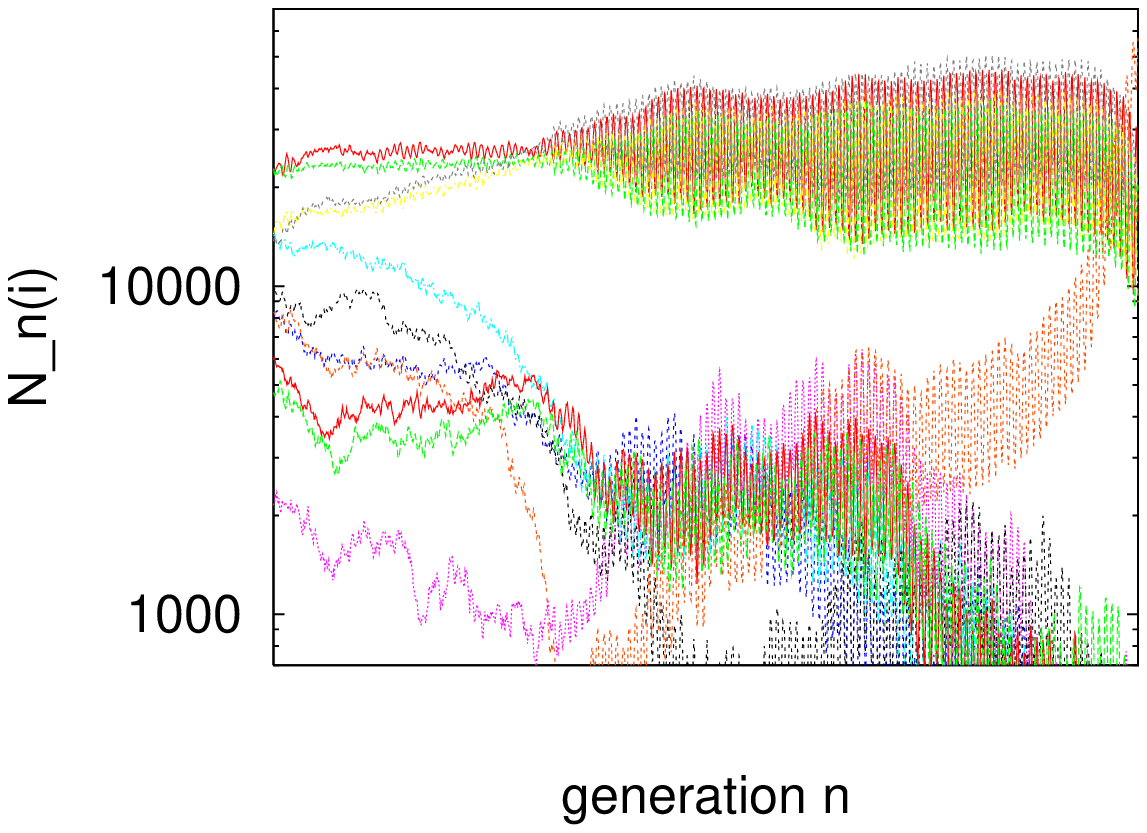,width=.53\textwidth}
\caption{The number of molecules $N_n(i)$ for the species $i$ is
plotted as a function of generation $n$ of cells, i.e., at each
successive division event $n$.  results from a random network with
$k=200$ and $\rho=.1$ was adopted, with $N=64000$ and $\mu=0.01$ (a),
and $\mu=0.1$ (b).  Only some species (whose population get large at
some generation) are plotted. in (a), a recursive production state is
established, while in (b), a few quasi-recursive states are
visited successively. (c): Expansion of Fig (b) around the time step 100000.}
\end{figure}

\begin{figure}
\noindent
(a)\epsfig{file=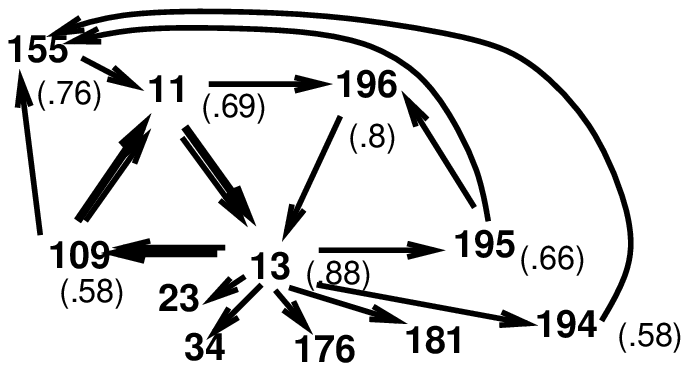,width=.5\textwidth}
(b)\epsfig{file=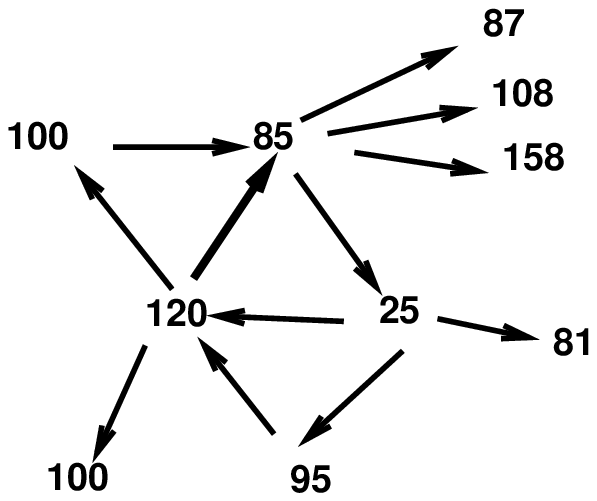,width=.4\textwidth}
\caption{The catalytic network of the dominant species that constitute
the recursive state. The catalytic reaction is plotted by an arrow $i
\rightarrow j$, as the replication of the species $j$ with the
catalytic species $i$.  The numbers in () denote $c_i$ of the species.
Only the species that continue to exist with the population larger
than 10 is plotted.  (Note many other species can exist at each
generation, through the replication error).  (a): corresponding to the
recursive state of Fig.9 a, where the three species connected by
thick arrows are the top 3 species in Fig.9 a.  The network (b) is
another example observed in a different set of simulations with
$k=200$ and $\rho =.1$, but with a different reaction network from
Fig.9.}
\end{figure}

In the phase (1), there is no clear recursive production and the
dominant molecule species changes by generation frequently. Even
though each generation has some dominating species as with regards to
the molecule numbers, the dominating species change every few
generations.  At one generation, some chemical species are dominant but
only a few generations later. Information regarding the previously
dominating species is totally lost often to the point that its
population drops to zero (see Fig.8).  Here no stable mutual catalytic
relationships are formed among molecules.  Hence, the time required
for reproduction of a cell is quite large, and much larger than the
case (2).

In the phase (2), a recursive state is established, and the chemical
composition is stabilized such that it is not altered much by the
division process (see Fig.9).  Generally, all the observed recursive states
consist of 5-12 species, except for those species with one or two
molecule numbers, which exist only as a result of replication errors.
These 5-12 chemicals mutually catalyze, by forming a catalytic network
as in Fig.10, which will be discussed later.  The member of these 5-12
species do not change by generations, and the chemical compositions
are transferred to the offspring cells.  Once reached, this state is
preserved throughout whole simulations, lasting over more than 10000
generations.

The recursive state observed here is not necessarily a fixed point
with regards to the population dynamics of the chemical
concentrations.  In some case, the chemical concentrations oscillate
in time, but the nature of the oscillation is not altered by the
process of cell division.


For example, in the recursive state depicted in Fig.9a), 11 species
remain in existence throughout the simulation. As shown, three species
have much higher populations than others, which form a hypercycle as
$109\rightarrow 11 \rightarrow 13 \rightarrow 109$. (The numbers
11,13,.. are indices of chemical species, initially assigned
arbitrarily).  The hypercycle sustains the replication of the
molecules, and is called 'core hypercycle'.  The catalytic activities
of the species satisfy $c_{13}>c_{109}>c_{11}$, and accordingly the
respective populations satisfy $N_{11} > N_{109} > N_{13}$.

In the phase (3), after one recursive state lasts over many
generations (typically a thousand generations), a fast switching state
appears until a new (quasi-)recursive state appears.  As shown in
Fig.9 b, for example, each (quasi-)recursive state is similar to that
in the phase (2), but in this case, its lifetime is finite, and it is
replaced by the fast switching state as in the phase (1).  Then the
same or different (quasi-)recursive state is reached again, which
lasts until the next switching occurs.  In the example of Fig.9b
(see also Fig.9c) for its expansion), around the 12000th generation,
the core network is taken over by parasites to enter the phase (1)
like fast switching state which in turn gives way for a new
quasi-recursive state around the 14000th generation.

In the example of Fig. 9b, there is another type of switching, as
shown around 85000th generation, as shown in Fig.9c with
magnification.  Here, the quasi-recursive state is still stable, but
the core hypercycle consisting of dominant species changes.  As in
Fig.9c, a switch occurs from an initial core hypercycle
($109$,$11$,$13$), to the next core hypercycle $(11,13,195,155)$
around the 8500th generation.

This latter switching is the competition among core networks, while
the former drastic switch is due to the invasion of parasitic
molecules, which is most commonly observed.  The mechanism of this
switching is discussed again in \S 5.2.4.

\subsubsection{Dependence of Phases on the Basic Parameters}

Although the behavior of the system depends on the choice of the
network, there is a general trend with regards to the phase change,
from (1), to (3), and then to (2) with the increase of $N$, or with
the decrease of $k$, as schematically shown in Fig.11.  By choosing a
variety of networks, however, we find a clear dependence of the
fraction of the networks on the parameters, leading to a rough sketch
of the phase diagram.  Generally, the fraction of (2) increases and
the fraction of (1) decreases also with the decrease of $\rho$ or $\mu$.
For example, the fraction of (1) (or (3)) gets
larger as $k$ is decreased from $k\stackrel{<}{\sim} 300$ for
$N=50000$ (with $\rho=.1$ and $\mu=.01$), while dependence on $\rho$
will be discussed below.

\begin{figure}
\noindent
\hspace{-.3in}
\epsfig{file=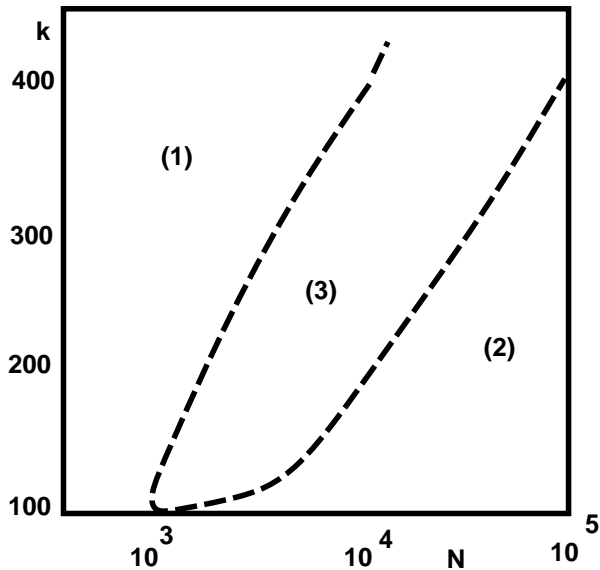,width=.5\textwidth}
\caption{Schematic representation of the phase diagram of the three
phases, plotted as a function of the total number of molecules $N$,
and the total possible number of molecule species $k$.}
\end{figure}

For a quantitative investigation, it is useful to classify the phases by
the similarity of the chemical compositions between two cell division
events\cite{Lancet}. To check the similarity, we first define a 
$k$-dimensional
vector $\stackrel{\rightarrow}{V_n}$=$(p_n(1),..,p_n(k))$ with $p_n(i)
=N_n(i)/N$. Then, we measure the similarity between $\ell$ successive
generations with the help of the inner product as

\begin{equation}
H_{\ell}=\stackrel{\rightarrow}{V_n} \cdot
\stackrel{\rightarrow}{V_{n+\ell}}/(|V_n||V_{n+\ell}|)
\end{equation}

In Fig.12, the average similarity $\overline{H_{20}}$ and the average
division time are plotted for 50 randomly chosen reaction networks as
a function of the path probability $\rho$.  Roughly speaking the
networks with $\overline{H_{20}}>.9$ belong to $(2)$, and those with
$\overline{H_{20}}<.4$ to $(1)$, empirically.  Hence, for $\rho >0.2$,
the phase (1) is observed for nearly all the networks (e.g. $48/50$),
while for lower path rates, the fraction of (2) or (3) increases. The
value $\rho \sim .2$ gives the phase boundary in this case.

Generally speaking, a positive correlation between the growth speed of
a cell and the similarity $H$ exists.  In Fig.12, the division time is
also plotted, where to each point with a high similarity $H$, a lower
division time corresponds.  The network with higher similarity (i.e.,
in the phase (2)) gives a higher growth speed.  Indeed, the recursive
states maintain higher growth speeds since they effectively suppress
parasitic molecules.  In Fig. 12, by decreasing path rates, the
variations in the division speeds of the networks become larger, and
some networks that reach recursive states have higher division speeds
than networks with larger $\rho$. On the other hand, when the path
rate is too low, the protocells generally cannot grow since the
probability to have mutually catalytic connections in the network is
nearly zero.  Indeed there exists an optimal path rate seems (e.g.,
around $.05$ for $k=200$, $N=12800$ as in Fig.12) for having a network
with high growth speeds. Consequently, under competition for growth,
protocells having such optimal networks will be evolved as will be
discussed in \S 5.3.

Besides the correlation between the growth speed and similarity, the
correlation with the diversity of the molecules also exists.
Protocells with higher growth speed and similarity in the phase (2)
have higher chemical diversity also.  In the phase (1), one (or a very
few) molecule species is dominant in the population, while about 10
species have higher population in the phase (2) with higher growth
speed, where the chemical diversity is maintained.

\begin{figure}
\noindent 
\epsfig{file=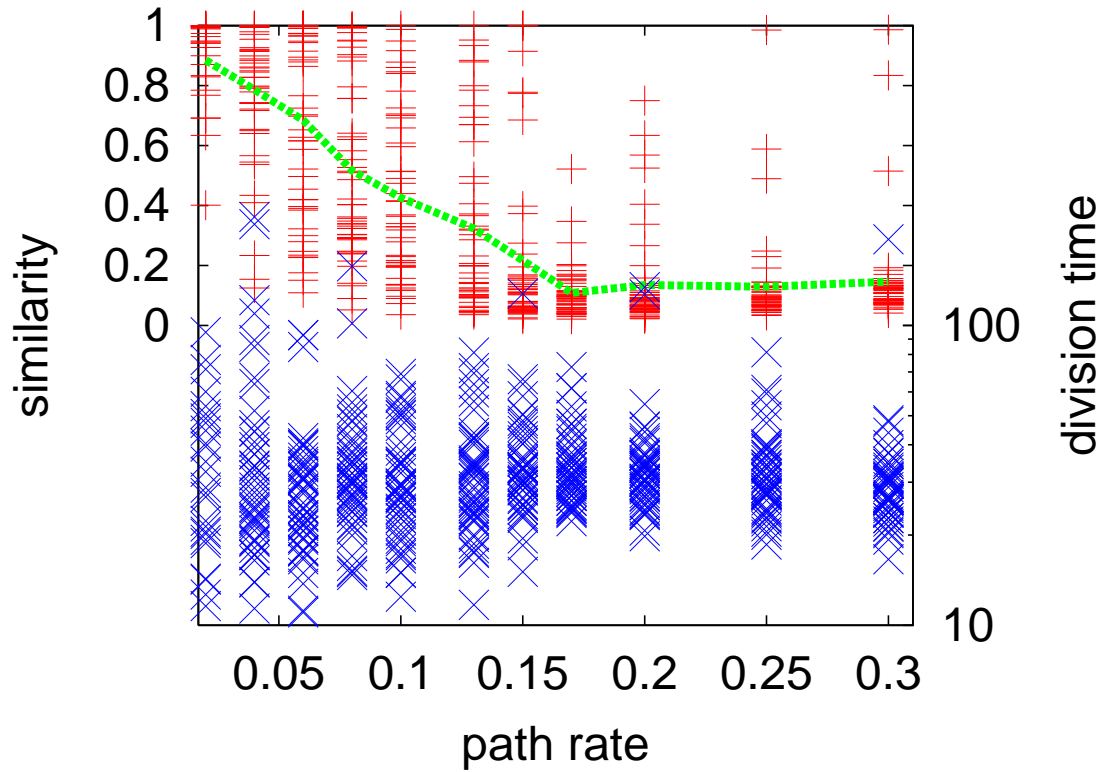,width=.95\textwidth}
\caption{The average similarity $\overline{H_{20}}$ ($+$), and the
average division time ($\times$) are plotted as a function of the path
rate $\rho$.  For each $\rho$, data from 50 randomly chosen networks
are plotted.  The average is taken over 600 division events.  The
dotted line indicates the average of $\overline{H_{20}}$ over the 50
networks for each $\rho$.
For $\rho>.2$, networks over 98 \% have $H<.4$, and they show fast switching,
while for $\rho=.08$, about 95\% belong to the phase (2) or (3)
At $\rho=0.02$, 25 out of 50 networks cannot support cell growth,
4 cannot at $\rho=0.04$. (Adapted from \cite{KK-PRE}).}
\end{figure}

\subsubsection{Maintenance of Recursive Production}

How is the recursive production sustained in the phase (2)?  We have
discussed already the danger of parasitic molecules that have lower
catalytic activities and are catalyzed by molecules with higher
catalytic activities. As discussed in \S 2.1, such parasitic molecules
can invade the hypercycle.  Indeed, under the structural changes and
fluctuations, the recursive production state could be destabilized.
To answer the question on the itinerancy and stability of
recursive states, we have examined several reaction networks.  The
unveiled logic for the maintenance of recursive state is summarized as
follows.

(a) {\bf Stabilization by intermingled hypercycle network}:

The 5-12 spices in the recursive state form a mutually catalytic
network, for example, as in Fig. 10.  This network has a {\sl core
hypercycle network}, as shown in thick arrows in Fig.10a.  As shown in
Fig.13, such core hypercycle has a mutually catalytic relationship,
as `` $A$ catalyzes $B$, $B$ catalyzes $C$, and $C$ catalyzes
$A$''. However, they are connected with other hypercycle networks such
as $G\rightarrow D \rightarrow B \rightarrow G$, and $D\rightarrow C
\rightarrow E \rightarrow D$, and so forth.  The hypercylces are
intermingled to form a network.  Coexistence of core hypercycle and
other attached hypercycles are common to the recursive states we have
found in our model.

This intermingled hypercycle network (IHN) leads to stability against
parasites and fluctuations.  Assume that there appears a parasitic molecule to one species in the
member of IHN (say $X$ as a parasite to $C$ in Fig.13).  The species
$X$ may decrease the number of the species $C$.  If there were only a
single hypercycle $A\rightarrow B \rightarrow C \rightarrow A$, the
population of all the members $A,B,C$ would be easily decreased by
this invasion of parasitic molecules, resulting in the collapse of the
hypercycle.  In the present case, however, other parts of the network
(say, that consisting of $A,B$,$G,D$ in Fig.13), compensate the
decrease of the population of $C$ by the parasite, so that the
population of $A$ and $B$ are not so much decreased.  Then, through
the catalysis of the species B, the replication of the molecule $C$
progresses, so that the population of $C$ is recovered.  Hence the
complexity in the hypercycle network leads to stability against the
attack of parasite molecules.

Next, IHN is also relevant to the stability against fluctuations.  It
is known that the population dynamics of a simple hypercycle often
leads to heteroclinic cycle\cite{Sigmund}, where the population of one
(or a few) member approaches 0, and then is recovered.  For a
continuum model, such heteroclinic cycle can continue forever, but in a stochastic
model, due to fluctuations, the number of the corresponding molecule
species is totally extinct sometimes.  Once this molecule species goes extinct
completely, and then its recovery by replication error would require a
very long time.  Hence, to achieve stability against fluctuations, a
state with the heteroclinic cycle dynamics or any oscillation in which
some of the population goes very low should be avoided.  Indeed, by forming IHN,
such oscillatory instability is often avoided or reduced.  Due to
coexistence of several hypercycle processes, instability in each
hypercycle cancels out, leading to fixed-point dynamics or oscillation
with a smaller amplitude.  Thus the danger that the population of some
molecules in the hypercycle goes to zero by fluctuations 
is reduced.

Stability of coexistence of many species is discussed as 'homeochaos'
\cite{homeochaos}, while stable reproduction in reaction network is
also seen in \cite{Ikegami}.

(b) {\bf Minority in the core hypercycle};

Now we study more closely the population dynamics in a core hypercycle.
Here, the number of molecules $N_j$ of molecule species $j$, is in the
inverse order of their catalytic activity $c(j)$, i.e,, $N_A>N_B>N_C$
for $c_A<c_B<c_C$.  Because a molecule with higher catalytic activity
helps the synthesis of others more, this inverse relationship is
expected. Indeed, the population sizes of just three species $A,B,C$,
with the catalytic relationship $A\rightarrow B \rightarrow C
\rightarrow A$ are estimated by taking the continuum limit $N
\rightarrow \infty$ and obtaining a fixed point solution of the rate
equation for the concentrations of the chemicals as discussed in
\cite{Eigen}.  From a straightforward calculation we have:
$N_A:N_B:N_C= c_A^{-1}:c_B^{-1}:c_C^{-1}$.

Here, the $C$ molecule is catalyzed by a molecule species with higher
activities but larger populations ($A$).  Hence, the parasitic
molecule species cannot easily invade to disrupt this mutually
catalytic network.  Since the minority molecule ($C$) is catalyzed by
the majority molecule ($A$) (with the aid of another molecule ($B$)),
a large fluctuation in molecule numbers is required to destroy this
network.

The stability in the minority molecule is also accelerated by the
complexity in IHN.  If the catalytic activity of $C$ is highest, the
recursive state here is mainly achieved by catalysis of the molecule
$C$.  On the other hand, this also implies that $C$ is the minority in
the core network.  (The population of the molecule $C$ is usually
larger than $D$,$E$, etc. in Fig.13, though.)  Hence the attack to $C$
molecule is most relevant to destroy this recursive state.  In the
IHN, this minority molecule species is involved in several hypercycles
as in $C$ in Fig.13.  This, on the one hand, demonstrates the
prediction in \S 4.5, that more species are catalyzed by the minority
molecules, while on the other hand, leads to the suppression of the
fluctuation in the number of minority molecules, as will be discussed
in \S 5.4.  With the decrease of the fluctuation, the probability that
the minority molecules is extinct is reduced, so that the recursive
state is hardly destroyed.

(c) {\bf Localization in a Random Network}

The present system belongs to a class of system with reaction and
diffusion, while the structural change by replication error leads to
the diffusion within the network space.  With random connection in the
catalytic network, the present system is nothing but a
reaction-diffusion in a random network.  Generally, such problem is
related with the Anderson localization, where concentrations are
localized within some part of the network, depending on the degree of
the connectivity in the network and the strength of the diffusion
coupling. From this viewpoint, the formation of IHN, localized only
within a limited species in the global network, may be understood as an
example of such localization.  It will be interesting to study the
stability of the recursive production, in terms of the localization
transition in the reaction network\cite{Takagi}.

\begin{figure}
\noindent
\epsfig{file=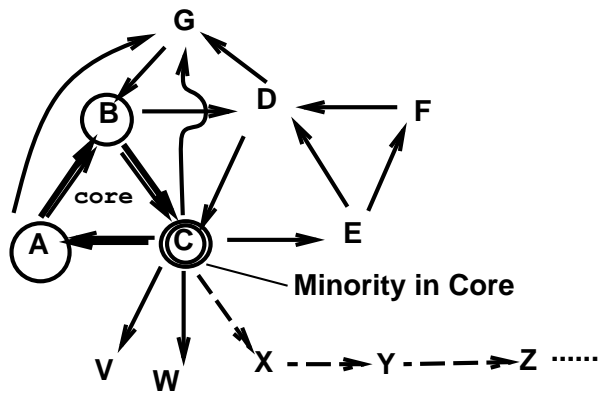,width=.5\textwidth}
\caption{An example of mutually catalytic network in our model.  The
core network for the recursive state is shown by circles, while
parasitic molecules ($X$,$Y$,..) connected by broken arrows, are
suppressed at a (quasi-)recursive state.}
\end{figure}

\subsubsection{Switching}

Next, we discuss the mechanism of switching.  In the phase (3), the
recursive production state is destabilized, when the population of
parasitic molecules increase.  For example, the number of the molecule
$C$ may be decreased due to fluctuations, while the number of some
parasitic molecules ($X$) that are not originally in the catalytic
network but are catalyzed by $C$, may increase.  Frequency of such
fluctuation increases as the total population of molecules in a cell
is smaller.  If such fluctuation appears, the other molecule species
in the original network loses the main source of molecules that
catalyze their synthesis, successively.  Then the new parasitic
molecule $X$ occupies a large portion of populations. However, the
molecule's main catalyst ($C$) soon disappears, the synthesis of $X$
is stopped, and this species $X$ is taken over by some molecules $Y$
that are catalyzed by $X$ (see the broken arrows in Fig.13).  Then,
within a few generations, dominant species changes, and recursive
production does not continue.  Indeed, this is what occurred in the phase
(1).  Then the parasitic molecule $X$ is taken over some other
$Y$. This take-over by parasites continues successively, until a new
(or same) recursive state with hypercycle network is formed.  Hence
the fluctuation in the minority molecule in the core network is
relevant to the switching process.

\subsection{Evolution}

{\bf Model A}

The next question we have to address is whether the recursive
production state is achieved through evolution.  To check this problem
we have extended our model to further include a ``mutational'' change
of network at each division event. (model A).
To be specific, at each division
event we add or delete randomly (with equal probability) a few
reaction paths, whose connection $i \rightarrow j$ is again chosen
randomly.  Here to see the evolution of catalytic activity, the index
of the species is ordered with the value of catalytic activity, i.e.,
the index $j$ is ordered so that $c_j$ monotonically increases with
$j$.  Since the mutational change is assumed to be random, a new path
is added or deleted independent of the catalytic activity.  In the
simulation displayed here, there are 5 mutations of the network path
at every generation.  We have carried out numerical experiments of this
model, to see if the path rate of the network stays around the state
supporting the recursive production.

\begin{figure}
\noindent
(a)\epsfig{file=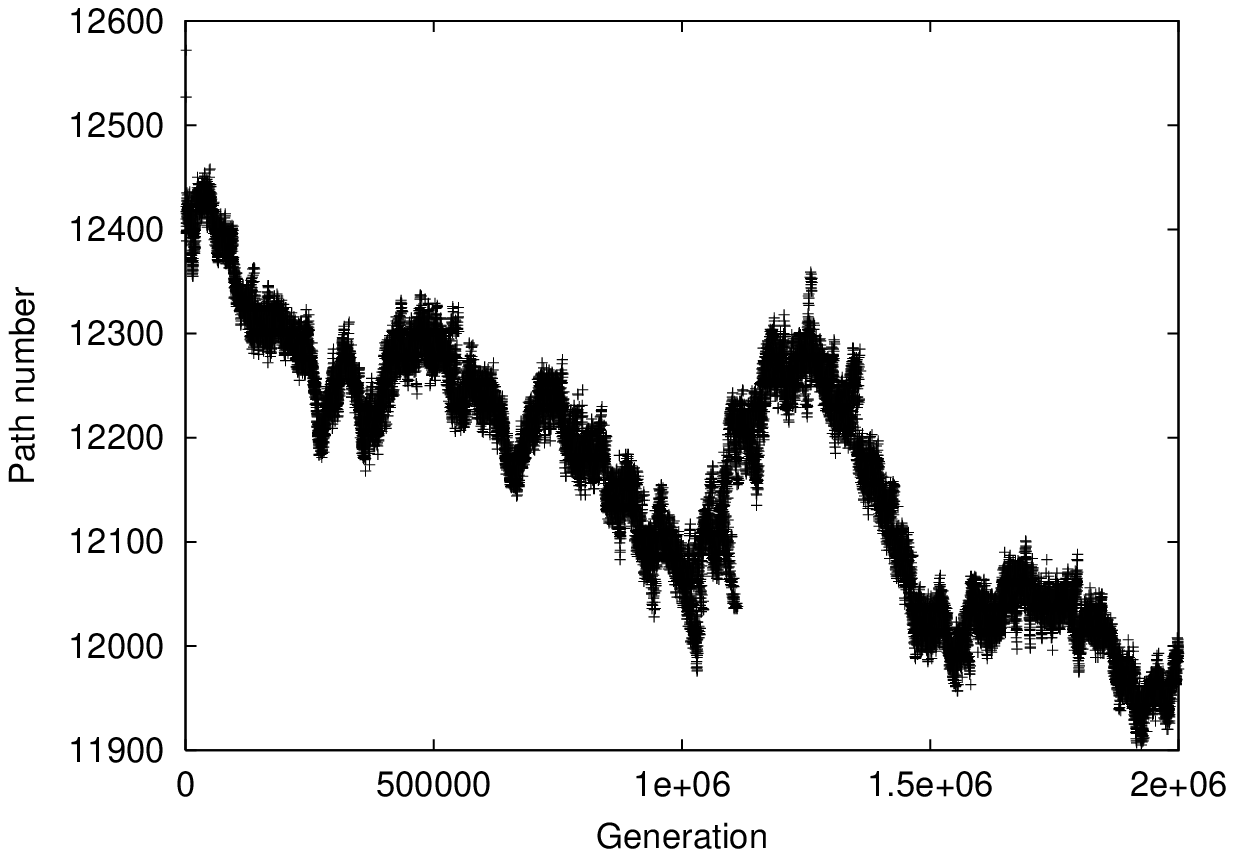,width=.55\textwidth}
(b)\epsfig{file=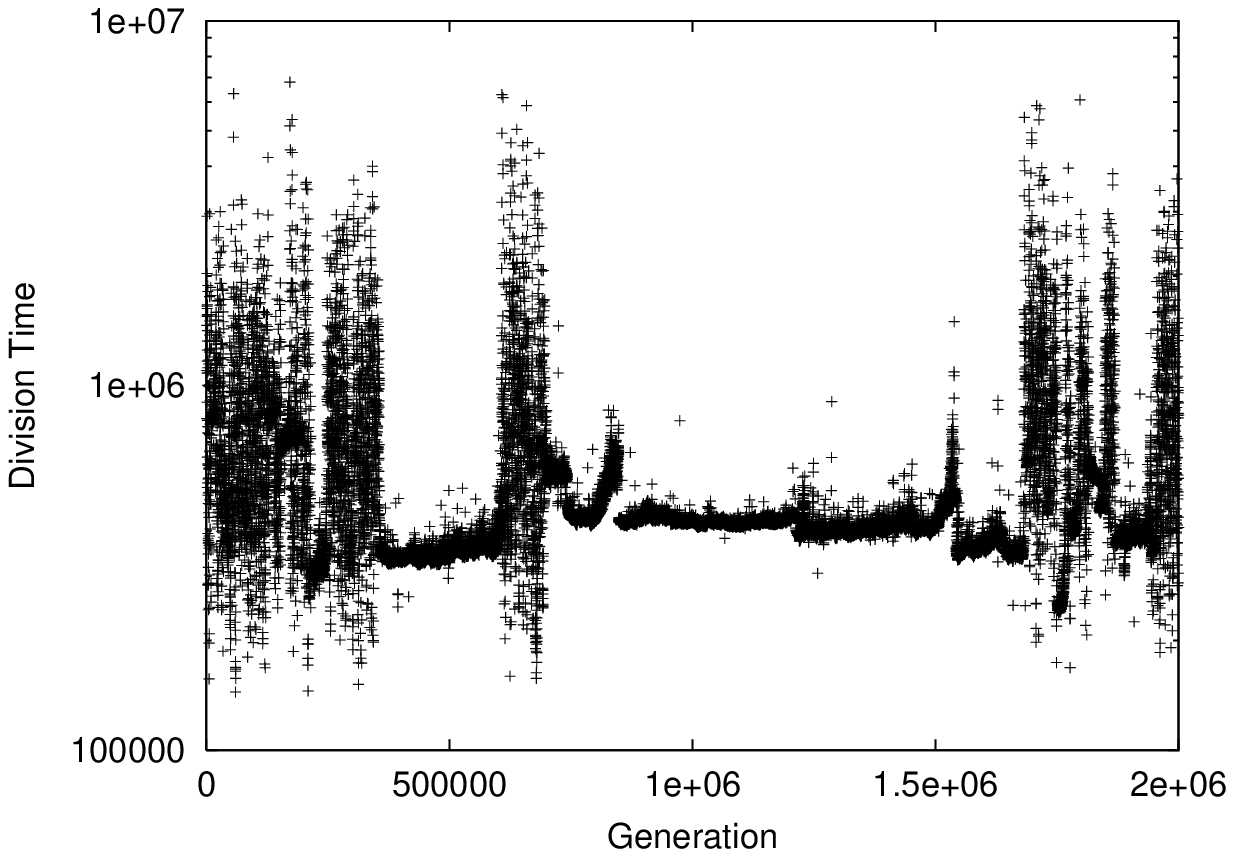,width=.55\textwidth}
(c)\epsfig{file=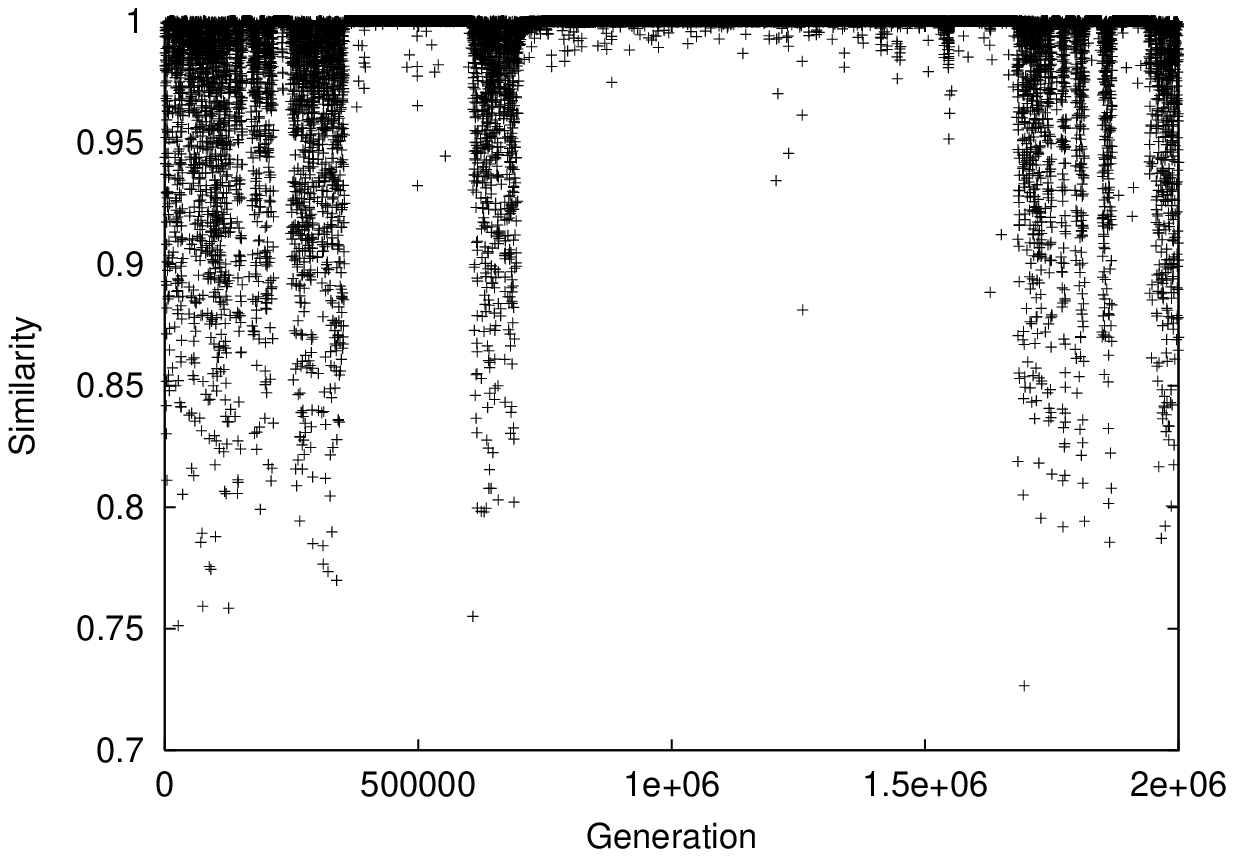,width=.55\textwidth}
\caption{Evolution of path-rates, recursiveness, and division time,
plotted versus generation.  The total number of species $k$ is 500,
where $c_i$ is chosen as $100^{-(k-i)/k}$, so that it ranges from
0.01 to 1.0 equally in logarithmic scale.  The number of molecules $N$ in
a cell is set at 50,000, so that the cell divided when the total
molecule number is 100,000. The initial path rate is set at
$\rho=0.1$, i.e., 125,000 paths totally.  At every division 5 paths
are "mutated", i.e., with equal probability 5 paths are added or
eliminated randomly.  Totally there are $M_{tot}=100$, so that one of
100 cells are eliminated when one cell is divided into two.  (a) the
total path number.  The path rate is obtained by dividing the number by
$k^2$.  (b) the division time, i.e., the required steps for a cell
divide (c) the similarity $H^1(i)$, defined in \S 5.2.}
\end{figure}


\begin{figure}
\noindent
\epsfig{file=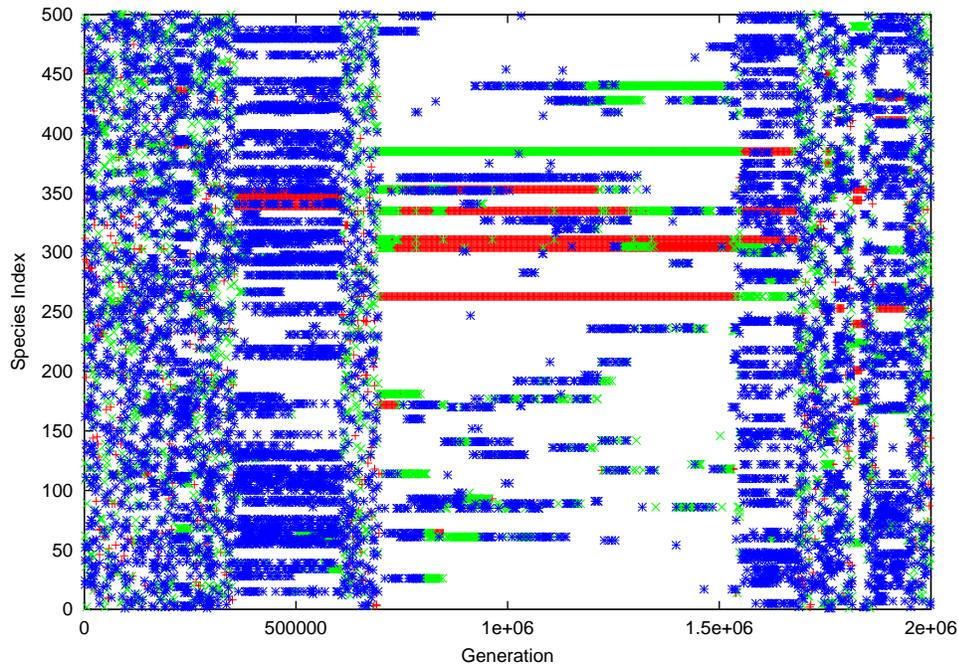,width=.8\textwidth}
\caption{Evolution of cell: Those species $i$ with $N(i)>100$ are
plotted with the vertical axis as the species index $i$, and the
longitudinal axis as the generation.  The data are from the result of
the simulation for Fig.14.}
\end{figure}

\begin{figure}
\noindent
\epsfig{file=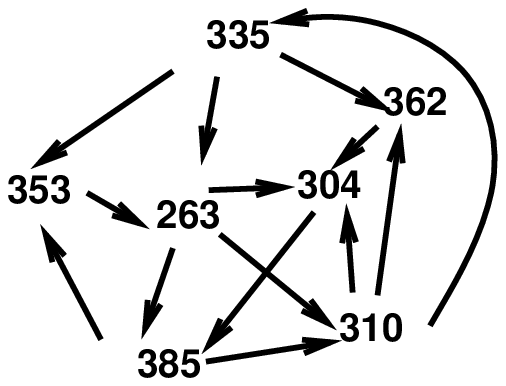,width=.5\textwidth}
\caption{The catalytic network of the species
that constitute the recursive state around $10^6$ th generation of Fig.14 or 15.
}
\end{figure}

An example of the time series of path rates at each generation is
shown in Fig.14, as well as the time series of the division time, and
chemical diversity.  Corresponding to this time series, the change of
dominant species is plotted over generations in Fig.15.

As shown, the recursive state is achieved, and is maintained over many
generations, until it switches to other states.  At each reproduction,
there are changes in the reaction paths here.  In spite of such
mutations, the recursive production state is sustained over many
generations.  In each recursive production state, the path rate
remains rather low.  Here, such network that supports the recursive
production is selected and is maintained. Note that many molecules are
catalyzed by the minority species in the core hypercycle network. In
this sense, a prototype of the evolution to package the information
into the minority molecule that is suggested in \S 4.5 is observed
here.

An example of the network of dominant species is given in Fig. 16.
Here intermingled hypercycle networks (IHN) are formed so that
recursive production is formed.  Again, there is a core hypercycle,
and other hypercycles are connected with it.  The surviving molecule
species have a large connectivity in reaction paths, much larger than
expected from a random network of the reaction path rate here.  As in
Fig 16. the IHN here forms a highly connected network, even though the
average path rate remains small (As shown in Fig.14, the path per
species is about 0.1 or lower).  The paths forming the IHN are
preserved over long generations, while a few paths are sometimes
eliminated.  Here, coexistence of several parallel paths among species
is important to give the robustness of the recursive state against
mutation that may delete one of the paths.  As in the dynamics of the
phase (3), the recursive production state is destabilized finally with
the mutation of reaction paths, while after some generations, other
recursive networks are formed through the mutation of the network.

To sum up, the phase (3) gives a basis for evolvability, since a
novel, (quasi-)recursive state with different chemical compositions is
visited successively.

{\bf Model B}

So far, we have assumed that the structural change in the replication
can occur equally to any other molecule species.  Of course, this is a
simplification, and the replication error occurs only to limited types
of molecules species that have similarity to the original.  To see
this point, we have  studied another model (model B) with some
modifications from the original model of \S 5.1.

Here, the catalytic activity is set as $c_i=i/k$, i.e., the activity
is monotonically increasing with the species index.  Then, instead of
global change to any molecule species by replication error, we modify
the rule so that the change occurs only within a given range $i_0 (\ll
k)$ i.e., when the molecule species $j$ is synthesized, with the error
rate $\mu$, the molecule $j+j'$ with $j'$ a random number over
$[-i_0,i_0]$ is synthesized.

In this {\bf model B}, we have not included any change of the network.
The network is fixed in the beginning, and is not changed through the
simulation.  Instead, by local change of structural error, the range
of species evolve by generations.  Here we take species only with
$i<i_{ini}$ in the initial condition, and examine if the evolution to
a network with higher catalytic activities (i.e., with much larger
$i$) progresses or not.  In other words, we examine if the indices $i$
in the network increase successively or not.  An example is shown in
Fig.17, where the catalytic activity increases through successively
switching to one (quasi-)recursive state ( consisting of species
within the width of the order $2i_0$ ), to another.

Here the switching occurs as in the phase (3). With the pressure for
selection of the protocells, cells with a new (quasi-)recursive state
are selected that consist of molecules with higher catalytic
activities (i.e., with larger indices of species).  Again each
recursive state consists of IHN, and the species with the highest
catalytic activity in the core hypercycle is minority in population.
Once the population of such species is decreased by fluctuations,
there occurs a switch to a new state that has higher catalytic
activities, and the species indices successively increase.  Hence,
evolution from a rather primitive cell consisting of low catalytic
activities to that with higher activities is possible, by taking
advantage of minority molecules.

Note that this switching cannot occur if the total number of molecules
$N$ is small.  When the number is too small, the mutation of paths to
destroy the recursive state hardly occurs.  On the other hand, if the
total number of molecules is too large, it is harder to establish a
recursive state, due to a larger possibility to change the network.
Hence, there is optimal value of the number of molecules in a
protocell to realize the recursive production as well as the
evolution.

\begin{figure}
\noindent
\epsfig{file=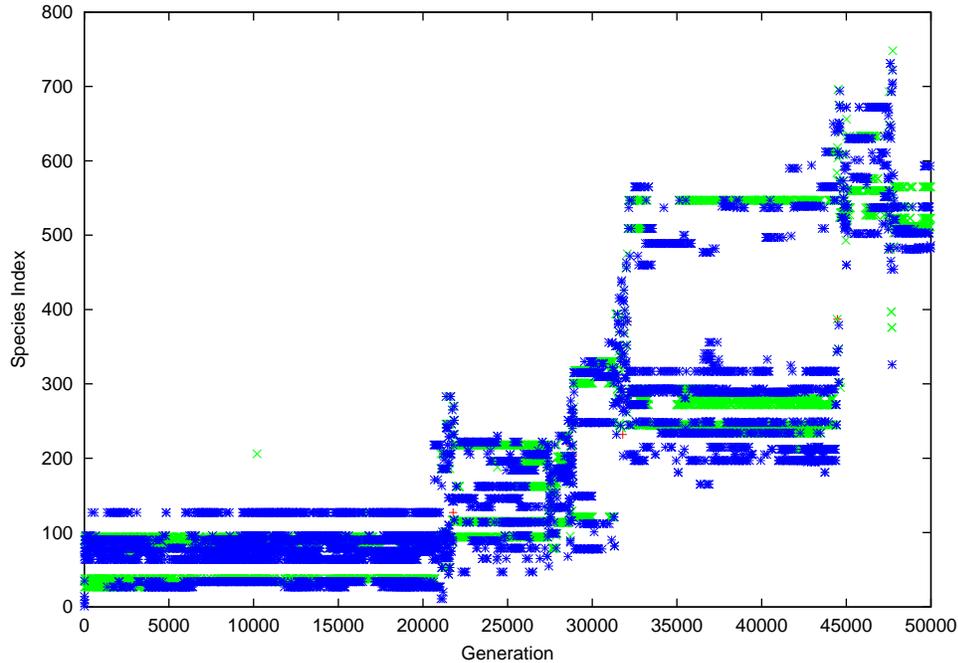,width=.8\textwidth}
\caption{Evolution of species in a cell:  Those species $i$ with $N(i)>100$ are plotted with the vertical axis
as the species index $i$, and the longitudinal axis as the generation.
The total number of species $k$ is 5000, where $c_i$ is chosen as
$c_i=i/k$, so that it ranges from 0.0002 to 1.0 equally distributed.
The number of molecules in a cell is set at 8,000, so that the cell divided
when the total molecule number is 16,000. The path rate is set at $\rho= 0.1$.
The replication error for the species occurs within the range of species
$[i-100,i+100]$,  instead of global selection from all species.
Totally there are $M_{tot}=10$ cells, 
so that one of 10 cells is eliminated when a cell is divided into two.}
\end{figure}

\subsection{Statistical Law}

To close the present section, we investigate the fluctuations of the
molecule numbers of each of the species, by coming back to the
original model studied in \S 5.2, without evolution of reaction paths.
The characteristics of the fluctuations of the number of each molecule
species over the generations can have a significant impact on the
recursive production of a cell, since the number of each molecule
species is not very large.  In order to quantitatively characterize
the sizes of these fluctuations, we have measured the distribution
$P(N_i)$ for each molecule species $i$, by sampling over division
events.

Our numerical results are summarized as follows:

(I) For the fast switching states, the distribution $P(N_i)$ satisfies
the power law

\begin{equation}
P(N_i) \approx N_i^{-\alpha},
\end{equation}

\noindent
with $1< \alpha \approx 2$, as shown in Fig. 18a.  The exponent $\alpha$ depends
on the parameters, and approaches 2 as alternation of dominant species is more frequent.
For example, as shown in Fig. 18b, thex exponent $\alpha$  increases from 1 to 2, with the
increase of the error rate $\mu$.

(II) For recursive states, the fluctuations in the core network
(i.e., 13,11,109 in Fig.9a or 10a) are typically small, (and are roughly
fit by Gaussian distribution).  On the other hand, for species that are
peripheral to but catalyzed by the core hypercycle, the number distribution is
closer to log-normal distributions

\begin{equation} P(N_i) \approx \exp(-\frac{(\log N_i-\overline{\log N_i})^2}{2\sigma}),
\end{equation}as shown in Fig.19.

Even though the distribution does not agree well with the log-normal distribution,
at least, the distribution if roughly symmetric after taking the logarithm
(i.e., as the 0-th approximation the distribution is not normal but log-normal).
The origin of the log-normal distributions here can be understood
by the following rough argument: for a replicating system, the
growth of the molecule number $N_m$ of the species $m$ is given by

\begin{equation}
dN_m/dt=AN_m,
\end{equation}

\noindent
where $A$ is the average effect of all the molecules that catalyze $m$.
We can then obtain the estimate

\begin{equation}
d\log N_m/dt =\overline{a} +\eta(t),
\end{equation}

\noindent
by replacing $A$ with its temporal average $\overline{a}$ plus
fluctuations $\eta(t)$ around it.  If $\eta(t)$ is approximated by a
Gaussian noise, the log-normal distribution for $P(N_m)$ is suggested
This argument is valid if $\overline{a}>0$.  As such this equation
diverges with time, but here, the cell divides into two before the
divergence becomes significant.  Although the asymptotic distribution
as $N \rightarrow \infty$ is not available then, the argument on the
distribution form is valid as long as $N$ is sufficiently large.

For the fast switching state, the growth of each molecule species is
close to zero on the average.  In this case the Langevin equation (12) can
approach 0, and we need to consider the equation by seriously
taking into account of the absorbing boundary condition at $N_m=0$.
By taking into account of the normalization of the probability, 
the stationary solution for the Fokker-Planck equation corresponding to eq.(12)
for $\overline{a} \leq 0$ is given by
\begin{equation}
P(N) \propto N^{-(1+\nu)},
\end{equation}
with
\begin{equation}
\nu =|\overline{a}|/(\overline{a^2}-\overline{a}^2).
\end{equation}
(see e.g., \cite{Sornette,Mikhailov-book}). Change of the exponent $\alpha$ against the
error rate in Fig.18b will be understood as the change of the ratio of variance 
to the mean of $a$. 

\begin{figure}
\noindent
(a)\epsfig{file=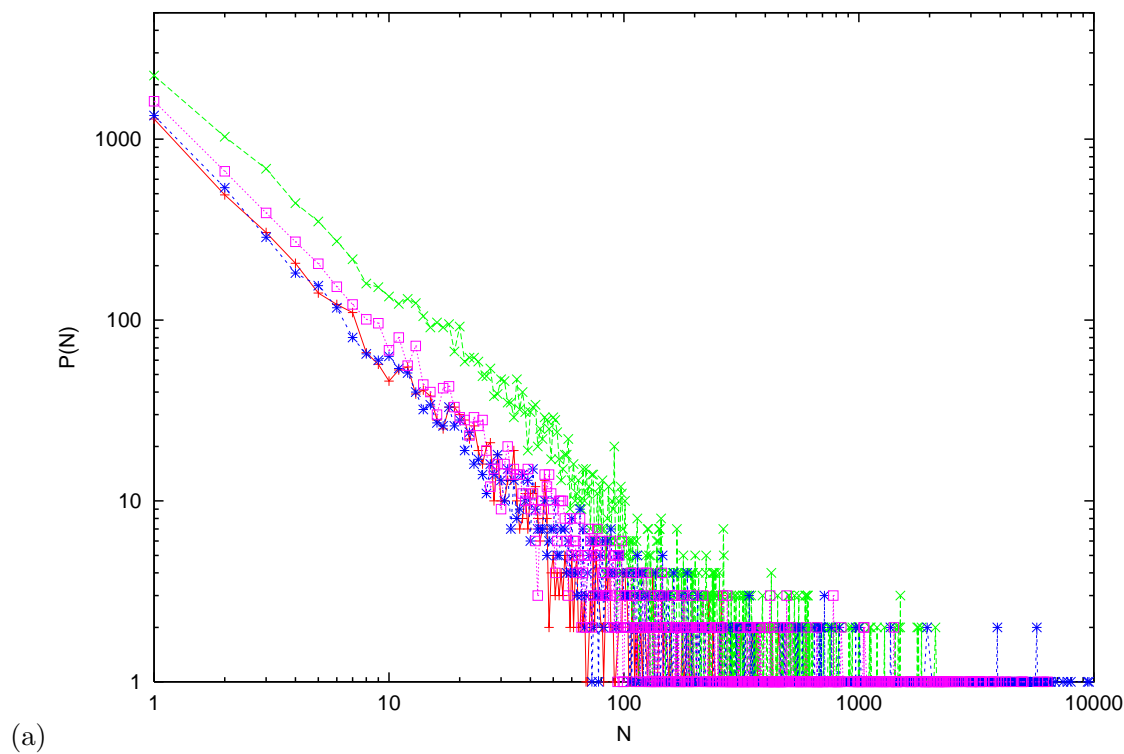,width=.9\textwidth}
(b)\epsfig{file=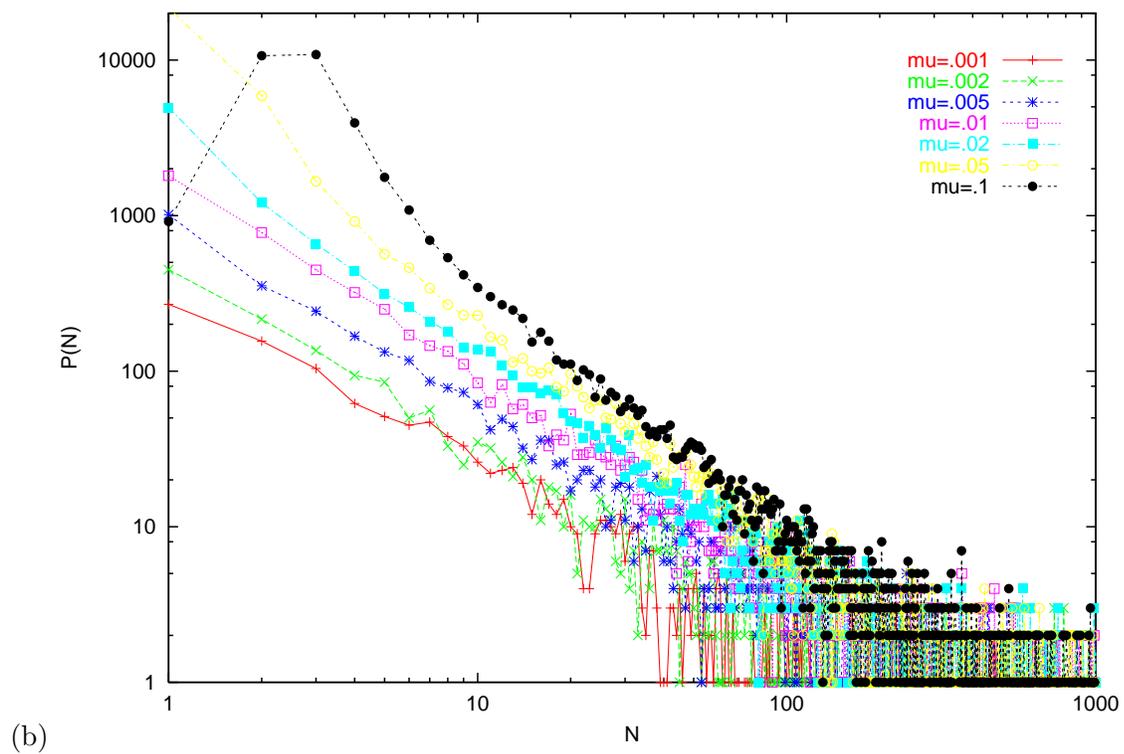,width=.9\textwidth}
\caption{The number distribution of the molecules corresponding to the
network in Fig.7 (fast switching states).  (a); The distribution is sampled
from 100000 division events. Plotted for 4 molecule species among 500.
Log-Log plot. (b) Change of the distribution with the change of the error rate $\mu$,
for a specific molecule species.}
\end{figure}

\begin{figure}
\noindent
\epsfig{file=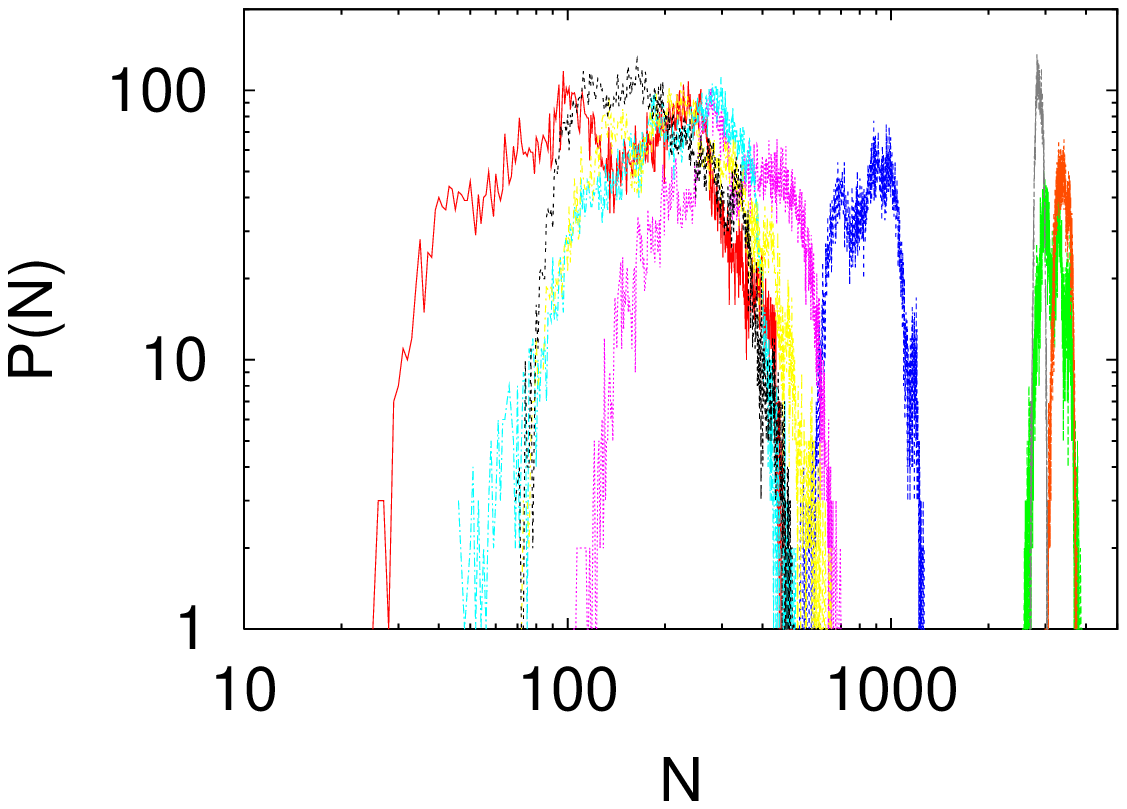,width=.9\textwidth}
\caption{The number distribution of the molecules corresponding to the
network in Fig.9a or 10a.  The distribution is sampled from 1000 division
events.  From right to left, the plotted species are
11,109,13,155,176,181,195,196,23.  Log-Log plot.}
\end{figure}

If several molecules mutually catalyze each other, however, one would
expect that the fluctuations will not increase as in the Brownian
motion as in eq. (12).  For example, consider that the number of one
species in the core cycle increase due to the fluctuation.  Then it
relatively decreases the number of molecules of the other species in
the core network, resulting in the suppression of the catalytic
reaction to replicate the increased species.  Then the catalytic
molecule of the original molecule species decreases.  Hence the
fluctuations in the core hypercycle is reduced.

Another reason for the reduction of fluctuation of the species in the
core cycle is high connectivity in the IHN.  The chemicals of core
part has catalytic paths with a large number of molecule species.
Hence many processes work in parallel to the synthesis of the core
species.  Then, fluctuations due to other chemical concentrations are
added in parallel.  Thus, the fluctuations can come close to Gaussian
distribution (recall the central limit theorem).

Note also that for some networks, the distributions of the molecule
numbers in the recursive sates may sometimes be intermediate between
log-normal and Gaussian, and occasionally even have double peaks.

By studying a variety of networks, the observed distributions of the
molecule numbers can be summarized as:

\begin{itemize}

\item
(1)Distribution close to Gaussian form, with relatively small
variances in the core (hypercycle) of the network.

\item
(2)Distribution close to log-normal, with larger fluctuations
for a peripheral part of the network.

\item
(3) Power-law distributions
for parasitic molecules that appear intermittently.

\end{itemize}

To quantitatively study the magnitude of variance in the IHN for the
recursive production, we have also plotted the variance
$\overline{(N_i-\overline{N_i})^2}$ ($\overline{..}$ is the average of
the distribution $P(N_i)$).  As can be seen in Fig.20, the variance in
the core network are small, especially for the minority species (i.e.,
13).  For molecule species that do not belong to the core hypercycle,
the variance scaled by the average increases as the average decreases.
Suppression of the relative fluctuation in the core hypercycle comes
from the direct feedback of the population change of the molecule
species in the core, as well as multiple parallel reaction paths, as
already mentioned.

\begin{figure}
\noindent
\epsfig{file=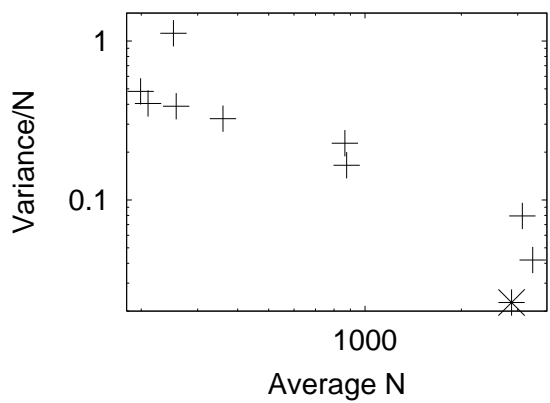,width=.5\textwidth}
\caption{Scaled variance, i.e., the variance of the molecule number
divided by its average is plotted against the average.  From the
largest to the smaller, the species 11 (the largest $\overline{N_i}$),
109(the second largest),13,155,194,176,195,181,196,23, 34(smallest
$\overline{N_i}$) are plotted.  Computed from the data in Fig.19.  The
asterisk denotes the species 13, that has largest catalytic activity here
and the minority in the hypercycle core. Adapted from \cite{KK-PRE}.}
\end{figure}

{\bf Remark: Universal Statistics}

Quite recently Furusawa and the author\cite{Zipf,
log} have studied several models of minimal cell consisting
of catalytic reaction networks, without assuming the replication
process itself.  In other words, the molecules are successively
synthesized from nutrition chemicals transported from the membrane,
where the level-(1) model of \S 3.2 is adopted.  They have found
universal statistical law of chemicals for a cell that grows
recursively.

(i) The number of molecules of each chemical species over all cells
generally obey the log-normal distribution.  This distribution is
universally observed for a state with recursive production.  Existence
of such log-normal distributions is also experimentally
verified\cite{Zipf}.  Ubiquity of log-normal distribution in
the level-(2) model described in this section is thus supported in the
level-(1) model.

(ii) A power law in the average abundances of chemicals.  This is
statistics against a huge number of molecule species. When the
abundances of all chemical species are ordered according to the
magnitude, the abundances of chemicals are inversely proportional to
the rank of the magnitude.  Such law was originally found in the
linguistics by Zipf\cite{Zipf-book}.  This Zipf's law on chemical abundances
\cite{Zipf} is found to be universal when a cell optimizes
the efficiency and faithfulness of self-reproduction. It is a
universal statistics when the cell model shows a recursive growth
under fluctuations in the molecule numbers.  Furthermore, using data
from gene expression databases on various organisms and tissues,
the abundances of expressed genes exhibit this
law.  Thus, the universal statistics are also supported
experimentally.  It is shown that this power law of gene expression is
maintained by a hierarchical organization of catalytic reactions.
Major chemical species are synthesized, catalyzed by chemicals with a
little less abundant chemicals.  The latter chemicals are synthesized
by chemicals with much less abundance, and this hierarchy of
catalytic reactions continues until it reaches the minor chemical
species.

{\bf Remark: Search for the deviation from universal statistics}

So far we have observed ubiquity of log-normal distribution, in
several models. The fluctuations in such distribution are generally
very large.  This is in contrast to our naive impression that a
process in a cell system must be well controlled.

Then, is there some relevance of such large fluctuations to biology?  Quite
recently, we have extended the idea of fluctuation-dissipation theorem
in statistical physics to evolution, and proposed linear relationship
(or high correlation) between (genetic) evolution speed and
(phenotypic) fluctuations.  This proposition turns out to be supported
by experimental data on the evolution of E Coli to enhance the
fluorescence in its proteins\cite{Sato}.  Hence the fluctuations are
quite important biologically.

The log-normal distribution is also rather universal in the present
cell, as demonstrated in the distribution of some proteins, measured
by the degree of fluorescence\cite{log}.  Now, is this universality the final
statement for "cell statistical mechanics"?  We have to be cautious
here, since too universal laws may not be so relevant to biological
function.  In fact, chemicals that obey the log-normal distribution
may have too large fluctuations to control some function.  Some other
mechanism to suppress the fluctuation may work in a cell.

Indeed, the minority control suggests the
possibility of such control to suppress the fluctuation, as discussed
in \S 4.5. For a recursive production system, some mechanism to
decrease the fluctuation in minority molecule may be evolved.

At least there can be two possibilities to decrease the fluctuation
leading to deviation from log-normal distribution.

The first one is some negative feedback process. In general, the
negative feedback can suppress the response as well
as the fluctuation.  Still, it is not a trivial question how chemical
reaction can give rise to suppression of fluctuation, since to realize
the negative feedback in chemical reaction, production of some
molecules is necessary, which may further add fluctuations.

The second possible mechanism is the use of multiple parallel reaction
paths.  If several processes work sequentially, the fluctuations would
generally be increased.  When reaction processes work in parallel for
some species, the population change of such molecule is influenced by
several fluctuation terms added in parallel.  If a synthesis (or
decomposition) of some chemical species is a result of the average of
these processes working in parallel, the fluctuation around this
average can be decreased by the law of large numbers.  Indeed, the
minority in the core network that has higher reaction paths has
relatively lower fluctuation as in Fig.20.  Suppression of fluctuation
by multiple parallel paths may be a strategy adopted in a cell.  Note
that this is also consistent with the scenario that more and more
molecules are related with the minority species as discussed in \S
4.5.  With the increase of the paths connected with the minority
molecules, the fluctuation of minority molecules is reduced, which
further reinforces the minority control mechanism.  Hence the increase
of the reaction paths connected with the minority molecule species
through evolution, decrease of the fluctuation in the population of
minority molecules, and enhancement of minority control reinforce
each other.  With this regards, search for molecules that deviate from
log-normal distribution should be important, in future.

In physics, we are often interested in some quantities that deviate
from Gaussian (normal) distribution, since the deviation is exceptional.  Indeed, in physics, search for
power-law distribution or log-normal distributions has been popular
over a few decades. On the other hand, a biological unit can grow and
reproduce, to increase the number.  For such system, the components
within have to be synthesized, so that amplification process is
common.  Then, the fluctuation is also amplified.  In such system, the
power-law or log-normal distributions are quite common, as already
discussed here, and as is also shown in several models and experiments
\cite{Zipf,log}.  In this case, the Gaussian (normal)
distribution is not so common (normal).  Then exceptional molecules
that obey the normal distribution with regards to their concentration
may be more important.

Also, the ubiquity of log-normal distribution we found is true for a
state with recursive production.  If a cell is not in a stationary
growth state but in a transient process switching from one steady
state to another, the universal statistics can be violated.  Search
for such violation will be important both experimentally and
theoretically.

\section{Summary}

We have studied a problem of recursive production and evolution  of a 
cell, by adopting a simple protocell system.  This protocell consists of
catalytic reaction network with replicating molecules.  The basic
concepts we have proposed through several simulations are as follows:

(i) {\bf Minority control}

In a cell system with mutually catalytic molecules, replicating
molecules with a smaller size in population are shown to control the
behavior of the total cell system.  This minority controlled state is
achieved by preserving rare fluctuations with regards to the molecule
number.  The molecule species, minority in its number, works as a
carrier of heredity, in the sense that it is preserved well with
suppressed number fluctuations and that it controls the behavior of a
cell relatively strongly.  Since molecules that are replicated by
this minority species are also preserved, more molecules will be synthesized
with the hep of it. In addition, 
reaction paths to stabilize the replication of this
minority molecules is expected to evolve.  Hence, the replication of more and
more molecule species is packaged into the synthesis of this minority
molecule, that also ensures the transmission of the minority molecule.
The minority molecule species, thus, gives a basis for "genetic
information".  Hence evolution from loose reproduction system to a
faithful replication system with genes is understood from a kinetic viewpoint
of chemical reaction.

(2) {\bf Recursiveness of production in an intermingled hypercycle network}

Next, a protocell model consisting of a variety of mutually catalyzing
molecule species is investigated.  When the numbers of molecules in a
cell is not too small and the number of possible species is not too
large in a cell, recursive production of a cell is achieved. This
recursive production state consists of 5-12 dominant molecule species,
which form intermingled hypercycle network(IHN).  Within this IHN, there
is a core hypercycle, while parallel multiple reaction paths in the
IHN are important to ensure the stability of the state against invasion
of parasitic molecules and against fluctuations in the molecule number.

(3) {\bf Itinerant dynamics over recursive production states}

When the fluctuation in molecule number is not small enough, there
appears switches over (quasi-)recursive production states.  A given
quasi-recursive state is destabilized by being taken over by some parasitic
molecules.  Then, the dominant molecule species change frequently by
generations, where the growth speed of a cell is suppressed.  After this
transient, the fast change of chemical compositions is reduced so that a
quasi-recursive production of a cell is sustained again.  Each switching
occurs with the loss of chemical diversity.  Note that in
high-dimensional dynamical systems, such switching over quasi-stable
states through unstable transient dynamics is studied as chaotic
itinerancy\cite{CI1,CI2,CI3}, where the loss of degrees of freedom is also
observed in the process of switching.

Destabilization of a recursive state in the present model occurs through
the decrease of the population of the minority molecules in the core
hyper cycle.  As this molecule species is taken over by parasitic
molecules, the switching starts to occur.  In this sense, the process in
the switching is not random, but is restricted to specific routes within
the phase space of chemical composition, as in the chaotic itinerancy.
It is interesting to study the present switching over recursive state as
a stochastic version of chaotic itinerancy.

(4){\bf Evolution through itinerant dynamics}

By considering change in the available reaction paths to the model, this
hypercycle network evolves to recursive production states.  Following the
itinerant dynamics above, each recursive state is later destabilized,
but later another recursive state is evolved. Through these successive
visits of recursive states, a cell can evolve to have a chemical
network supporting a higher growth speed.  Since the minority species in
the hypercycle network is relevant to this switch, minority molecules
are shown to be important to evolution.

(5){\bf Universal statistics and control of fluctuations}

Statistics of the number fluctuations of each molecule species is
studied. We have found that (i)power-law distribution of fast switching
molecules (ii) suppression of fluctuation in the core hypercycle species
and (iii) ubiquity of log-normal distribution for most other molecule
species.  The origin of log-normal distribution is generally due to
multiplicative stochastic process in the catalytic reaction dynamics,
as is confirmed in several other reaction network models.  On the other
hand, suppression of the number fluctuations of the core hypercycle is
due to high connections in reaction paths with other molecules.  In
particular, reduced is the number fluctuations of the minority molecule
species that has high catalytic connections with others.  This
suppression of fluctuation further reinforces the minority control for
the reproduction of a cell.  The deviation from ubiquitous log-normal
distribution thus appears, which may be important in control of cell
function.

In the present paper, we have not discussed cell-cell interaction, and
restricted our study only to a production process of a single cell.  Of
course, cells start to interact with each other, as the cell density is
increased through the cell division.  Indeed, including the cell-cell
interaction to the present cell model with reaction network, cell
differentiation and morphogenesis of a cell aggregate are
studied\cite{KKTY,Furusawa}.  Through instability of
intra-cellular dynamics with cell-cell interaction, cell
differentiation, irreversible loss of plasticity in cells, and robust
pattern formation process appear as a general course of development with
the increase of the cell number.  Relevance of minority control and
deviation from universal statistics to such multicellular developmental
process will be an important issue to be studied in future.

{\sl acknowledgments}

The author is grateful to T. Yomo, C. Furusawa, W. Fontana, Y. Togashi,
A. Awazu, and K. Fujimoto for discussions.  The work is partially
supported by Grant-in-Aids for Scientific Research from the Ministry of
Education, Science, and Culture of Japan (11CE2006).


\begin{thebibliography}{999}

\bibitem{whatlife}
K. Kaneko 'What is Life?: A complex systems approach", in Japanese,
Univ Tokyo Press. 2003

\bibitem{minority}
K. Kaneko, T. Yomo, 
J. Theor. Biol. 214 (2002) 563-576

\bibitem{Shannon}
C. Shannon and W. Weaver ``The Mathematical Theory of Communication",
Univ. of llinois Press, 1949

\bibitem{Brillouin}
L. Brillouin,
{\sl Science and Information Theory},
Academic Press 1969

\bibitem{Barabasi}
H. Jeong, et al., {\it Nature} {\bf 407}, 651 (2000);
H. Jeong, S. P. Mason, A.-L. Barab\'{a}si, {\it Nature} {\bf 411}, 41 (2001).

\bibitem{Spiegelman}
D.R. Mills, R.L. Peterson, and S. Spiegelman,
Proc. Nat. Acad. Sci. USA 58 (1967) 217;
D.R. Mills, F.R. Kramer, and S. Spiegelman,
Science 180 (1973) 916

\bibitem{Eigen}
M. Eigen and  P. Schuster, {\sl The Hypercycle} (Springer, 1979).

\bibitem{Hogeweg}
M. Boerlijst and P. Hogeweg, Physica 48D (1991) 17;
P.Hogeweg
Physica 75 D (1994)275-291

\bibitem{Dyson}
F. Dyson, {\sl Origins of Life}, Cambridge Univ. Press., 1985

\bibitem{Kauffman}
S.A. Kauffman, {\sl The Origin of Order}, Oxford Univ. Press. 1993

\bibitem{Bagley}
R.Bagley, J.D. Farmer, S. Kauffmans,
pp 93-140, in {\sl Artificial Life} 1989, ed. C. Langton

\bibitem{Cairns-Smith}
A.G. Cairns-Smith,
Clay Minerals and the Origin of Life(1982), Cambridge Univ. Press.

\bibitem{mtb}
K. Kaneko, 
in {\sl Function and Regulation of Cellular Systems} (2003)
Birkhauser (ed. A. Deutsch et al.)

\bibitem{Complexity}
K. Kaneko,
Complexity, 3 (1998c) 53-60

\bibitem{KKTY}
K. Kaneko and T. Yomo, 
Physica 75 D (1994), 89-102;
B. Math.Biol. 59 (1997) 139;
J. Theor. Biol., 199 243-256 (1999)

\bibitem{Furusawa}
Furusawa C. \& Kaneko K.,
Bull.Math.Biol. 60(1998) 659-687;
Phys Rev Lett. 84:6130-6133
J. Theor. Biol.  209 (2001) 395-416;
Anatomical Record, 268 (2002) 327-342;
J. Theor. Biol. 224 (2003) 413-435.

\bibitem{speciation}
K. Kaneko and T. Yomo,
Proc. Roy. Soc. B, 267 (2000) 2367-2373;
K. Kaneko, 
Population Ecology, 44 (2002) 71-85

\bibitem{Matsuura}
T. Matsuura, T. Yomo, M. Yamaguchi, N. Shibuya., E.P. Ko-Mitamura, Y. Shima, and
I. Urabe
Proc. Nat. Acad. Sci. USA  99 (2002) 7514-7517

\bibitem{Ko}
E. Ko, T.Yomo, and I. Urabe, Physica 75 D (1993)81-88

\bibitem{Kashiwagi1}
Kashiwagi A., Noumachi W., Katsuno M., Alam M.T., Urabe I., and Yomo T.
J. Mol. Evol., (2001)  {\bf 52}  502-509 .

\bibitem{Kashiwagi2}
A. Kashiwagi, I. Urabe, K. Kaneko, T. Yomo, submitted (2003)

\bibitem{Asashima}
T. Ariizumi and M. Asashima,
Int. J. Devl Biol. 45 (2001) 273-279


\bibitem{AL}
C. Langton eds.  Artificial Life 1989, Adisson Wesley

\bibitem{Fontana}
W. Fontana and L.W. Buss, 1994. 
Bull Math Biol 56:1-64

\bibitem{Awazu}
A. Awazu and K. Kaneko, preprint 2003.

\bibitem{Zipf}
C. Furusawa and K. Kaneko, Phys. Rev. Lett. 90 (2003) 088102.

\bibitem{Cell}B. Alberts, D.Bray, J. Lewis, M. Raff, K. Roberts, and J.D. Watson,
{\sl The Molecular Biology of the Cell}, 1983,1989,1994,2002

\bibitem{Mikhailov}
B. Hess and A. Mikhailov,
Science {\bf 264}, 223 (1994);
A. Mikhailov and B. Hess,
J. Theor. Biol. {\bf 176}, 185-192 (1995).

\bibitem{Togashi}
Y. Togashi and K. Kaneko,
Phys. Rev. Lett. , 86 (2001) 2459;
J.Phys.Soc.Japan 72 (2003)62-68;
preprint 2003.

\bibitem{Szathmary}
E. Szathmary and J. Maynard Smith,
J. Theor. Biol. 187 (1997) 555-571

\bibitem{Eigen-book}
M. Eigen,  Steps towards Life, Oxford Univ. Press., 1992

\bibitem{KK-net}
K. Kaneko, J. Biol. Phys., 28 (2002) 781;
Adv. in Complex Systems, 6 (2003)79-92

\bibitem{KK-PRE}
K. Kaneko, Phys. Rev.E. 68 (2003) 031909;

\bibitem{Lancet}
D. Segr\'{e}, D. Ben-Eli, D. Lancet,
Proc. Natl. Acad. Sci. USA 97 (2000)4112;
D. Segr\'{e} et al., J. theor. Biol. {\bf 213} (2001) 481
D. Segr\'{e} and D. Lancet, EMBO Reports {\bf 1} (2000) 217,

\bibitem{Sigmund}
J. Hofbauer and K. Sigmund,
{\sl Evolutionary Games and Population Dynamics},
Cambridge Univ. Press. 1998

\bibitem{homeochaos}
K. Kaneko and T. Ikegami, 
Physica D 56 (1992) 406-429

\bibitem{Ikegami}
T. Ikegami and T. Hashimoto,
Artificial Life 2 (1996) 305-318

\bibitem{Takagi}
H. Takagi and K. Kaneko, preprint (2003)

\bibitem{Mikhailov-book}
A. S. Mikhailov \& V. Calenbuhr, ``From Cells to Societies''
Springer 2002

\bibitem{Sornette}
D.Sornette, {\sl Critical phenomena in Natural Science}, Springer 2002

\bibitem{Zipf-book}
G. K. Zipf, {\it Human Behavior and the Principle of Least Effort}
(Addison-Wesley, Cambridge, 1949).


\bibitem{log}
C. Furusawa, T. Suzuki, A. Kashiwagi, T. Yomo and K. Kaneko ;
Ubiquity of Log-normal Distribution in gene expression,
preprint

\bibitem{Sato}
K. Sato, Y. Ito, T. Yomo, and K. Kaneko;
Proc. Nat. Acad. Sci. USA 100 (2003) 14086-14090

\bibitem{CI1}  
K. Kaneko, 
Physica D 41(1990) 137-172

\bibitem{CI2}
I. Tsuda, Neural Networks 5(1992)313

\bibitem{CI3}
K. Kaneko and I. Tsuda. ed.,  Focus issue on  ``Chaotic Itinerancy",
Chaos. 13 (2003) 926

\end{thebibliography}
\end{document}